\documentclass[aps,prd,preprint,eqsecnum,tightenlines,floats,
groupedaddress,showpacs]{revtex4}
\usepackage{amssymb} 
\usepackage{graphicx}
\usepackage{subfigure}
\usepackage{dcolumn}

\begin{document}


\date{June 8, 2021}

\title{Implications of triangular features in the Gaia skymap 
for the Caustic Ring Model of the Milky Way halo}

\author{Sankha S. Chakrabarty$^{a,b,c}$, Yaqi Han$^c$,
Anthony H. Gonzalez$^d$, and Pierre Sikivie$^c$\\
\vskip 0.25cm
$^a$ {\it Dipartimento di Fisica, Universit\`a di Torino,  
Via P. Giuria 1, I-10125 Torino, Italy}\\
$^b$ {\it Istituto Nazionale di Fisica Nucleare (INFN), 
Sezione di Torino, Via P. Giuria 1, I-10125 Torino, Italy}\\
$^c$ {\it Department of Physics, University of Florida,
Gainesville, FL 32611, USA}\\
$^d$ {\it Department of Astronomy, University of Florida, 
Gainesville, FL 32611, USA}}

\vskip 1cm

\begin{abstract}

The Gaia map of the Milky Way reveals a pair of triangular 
features at nearly symmetric locations on opposite sides 
of the Galactic Center.  In this paper we explore the 
implications of these features assuming they are manifestations 
of a caustic ring in the dark matter distribution of the Milky 
Way halo.  The existence of a series of such rings is predicted 
by the Caustic Ring Model.  The model's phase-space distribution 
is that acquired by a rethermalizing Bose-Einstein condensate 
of axions or axion-like particles.  We show that dust is 
gravitationally entrained by cold axion flows and propose 
this as an explanation for the sharpness of the triangular 
features. The locations of the features imply that we on 
Earth are much closer to the fifth caustic ring than thought 
on the basis of pre-Gaia observations.  Most likely we are 
inside its tricusp cross-section.  In that case the dark 
matter density on Earth is dominated by four cold flows, 
termed Big, Little, Up and Down.  If we are outside the 
tricusp cross-section the dark matter density on Earth is 
dominated by two cold flows, Big and Little.  We use the 
triangular features in the Gaia map, and a matching feature 
in the IRAS map, to estimate the velocity vectors and 
densities of the four locally dominant flows.

\end{abstract}

\maketitle

\section{Introduction}

The identity of dark matter is widely regarded as one of the most 
central and important questions in science today.  A related 
question is: what is the phase-space distribution of dark matter 
in galactic halos, and particularly what is the phase-space 
distribution of dark matter in the halo of our own Milky Way 
galaxy.  This latter question is important for at least two 
reasons.  First, the Milky Way halo is the environment in 
which the Galactic Disk forms and lights up with stars such 
as our Sun.  Second, knowledge of the phase-space distribution 
helps direct and indirect dark matter searches on Earth.  
Finally, there is the tantalizing possibility that the 
phase-space distribution depends on the identity of the 
dark matter particle, and hence that observations of the 
phase-space distribution may reveal or constrain the dark 
matter's identity.   

Dark matter is generally thought to be cold and collisionless.
``Cold" means that the primordial velocity dispersion of the 
dark matter particles is small, less than of order $10^{-8}$ 
today. Throughout, we use units in which $\hbar = c = k_B = 1.$   
``Collisionless" means that, to excellent approximation,  
gravity is the only relevant interaction among the dark 
matter particles as far as large scale structure formation 
is concerned.  Particle candidates with these properties 
include weakly interacting massive particles (WIMPs), QCD 
axions or axion-like particles (ALPs), and sterile neutrinos
\cite{PDM}.

Cold collisionless dark matter lies on a thin 3-dimensional 
hypersurface in 6-dimensional phase-space \cite{Ipser}.  We 
refer to this 3-dimensional hypersurface as the phase-space 
sheet. Its qualitative behavior is described in Fig. 1. The 
thickness of the phase-space sheet is the primordial velocity 
dispersion.  In the non-linear regime of structure formation, 
the phase-space sheet wraps around massive objects, such as 
galaxies, in phase-space.  As a result, at any physical point 
inside a galactic halo there is an odd number flows.  Furthermore, 
the dark matter density diverges in the limit of zero velocity 
dispersion on the surfaces in physical space across which the 
number of flows changes by two.  Such surfaces are called 
caustics.  The number of flows expected on Earth is of order 
one hundred \cite{Ipser}.  Present N-body simulations have 
inadequate resolution for revealing any but a small number 
of the expected flows.  Arguments for the robustness of 
cold flows and caustics in galactic halos are given in 
ref.~\cite{robust}.

The Caustic Ring Model is a proposal for the phase-space 
distribution of the dark matter halo of the Milky Way, 
and those of other disk galaxies \cite{Duffy}.  The model 
is characterized by axial symmetry, self-similar time 
evolution,  and large scale vorticity.  Although real 
galaxies are not exactly axial symmetric, they are 
sufficiently so that the model may apply to them
as a first approximation.  According to the model, 
the inner caustics in the halos of disk galaxies are 
rings.  Caustic rings are closed tubes whose cross-section 
is described by the elliptic umbilic ($D_{-4}$) catastrophe 
\cite{sing}.  The cross-section has the shape of a tricusp; 
see Fig. 2.  The rings lie in the galactic plane.  Their 
radii in any given disk galaxy are predicted in terms of 
a single adjustable parameter $j_{\rm max}$; see Eq.~(\ref{crr}).  
Observational evidence is claimed in support of the model.

Large scale vorticity is not possible if the dark matter is 
WIMPs or sterile neutrinos \cite{Natarajan}.  QCD axions 
\cite{PQWW,invisible} and ALPs \cite{Arias} produced by 
the vacuum realignment mechanism \cite{axdm}, including 
ultra-light ALPs, differ from WIMPs and sterile neutrinos in 
that they form a Bose-Einstein condensate \cite{CABEC,Erken,
Christopherson}.  Thermalization is a necessary condition for 
Bose-Einstein condensation.  The interaction through which 
cold dark matter axions thermalize is self-gravity.  When 
falling onto galactic halos, cold dark matter axions 
thermalize sufficiently quickly that they almost all go 
to the lowest energy state available to them consistent 
with the total angular momentum they have acquired through 
tidal torquing \cite{Erken,angmom}.  That state is one of 
rigid rotation in the angular variables.  Thus axion (or ALP) 
dark matter does acquire vorticity in galactic halos, and 
accounts for the existence of caustic rings.  It was shown 
that axion Bose-Einstein condensation accounts in fact for 
all properties of the Caustic Ring Model \cite{case}.  The 
observational evidence for caustic rings and the Caustic 
Ring Model is therefore evidence that the dark matter is 
a Bose-Einstein condensate, of axions or ALPs, at least 
in part.

The idea that the cold axion fluid thermalizes by 
gravitational self-interactoions, and as a result 
forms a Bose-Einstein condensate, was criticized 
in refs. \cite{Davidson1,Davidson2,Guth}.  These 
criticisms were addressed in ref. \cite{SC}.

One piece of evidence that has been claimed in support of 
caustic rings is a triangular feature in the IRAS map of 
the Milky Way plane; see 
\url{https://www.phys.ufl.edu/~sikivie/triangle/} .
The triangular feature is interpreted as the imprint
of the 5th caustic ring of dark matter in our galaxy 
upon baryonic matter in the Galactic Disk, seen in 
a direction tangent to the ring from our perspective
\cite{MWcr}.  The IRAS triangle is in the direction 
of Galactic Coordinates $(l,b) = (80^\circ, 0^\circ)$.  
The IRAS map does not have a matching triangular feature on 
the other side, near $(l,b) = (-80^\circ, 0^\circ)$.  However, 
the recently released Gaia skymap \cite{Gaia1,Gaia2} has two 
triangular features.  One of these coincides with the IRAS 
triangle on the left side.  The other is on the right 
side, at $(l,b) = (-91^\circ, 0^\circ)$; see
\url{https://www.phys.ufl.edu/~sikivie/Gaiamap/} .
Our paper assumes the two features are effects of the 
5th caustic ring on baryonic matter and explores the 
implications of this for the Caustic Ring Model.  The 
fact that the two triangles do not match perfectly in 
direction is attributed to a displacement of the caustic 
ring center $5.8^\circ$ towards the right relative to 
the Galactic Center.

Section III of our paper is concerned with the mechanism by 
which the caustic ring produces triangular features in the 
IRAS and Gaia maps.   It was proposed in ref.~\cite{MWcr} 
that the features are produced by gas and dust in thermal 
equilibrium in the gravitational field of the caustic ring.  
However, the features produced in this way have been found to 
be much less sharp than the observed features \cite{starsgas}.  
We propose here instead that dust is entrained by the cold 
axion flows forming the nearby caustic.  Because the dust 
particles follow the same trajectories as the axions, they 
form the same caustics.  This proposal accounts for the 
sharpness of the triangular features.  We derive a formula 
for the drag on a dust particle moving with respect to 
a cold axion fluid and compare this to the drag on the 
dust particle as a result of collisions with gas in the 
disk.  A major source of uncertainty is the temperature
of the axions.  We discuss the present axion temperature 
considering primordial heat in the axions themselves and 
heat absorbed later by cooling baryons

When only the IRAS triangle was known, it was thought
that the Sun lies close to, but outside of, the tricusp 
volume of the nearby caustic ring because it was assumed
that the caustic ring center coincides with the Galactic 
Center.  Now, taking account of the Gaia right triangle, 
the Sun is much closer to the caustic than previously 
thought.  In fact we believe the Sun almost certainly
is inside the tricusp volume of the 5th caustic ring.  
There are then four prominent flows on Earth associated 
with that caustic.   Our goal in Section IV is to derive 
the properties of the four flows from the observed 
triangular features.

The outline of our paper is as follows.   Section II 
describes caustic rings, the Caustic Ring Model and the 
observational evidence in support of it.  The rotation 
curve of M31 and the Gaia triangles are presented as 
additional evidence.  Section III discusses the 
entrainment of dust by cold axion flows, and the 
effect of dust-gas and dust-dust collisions on the 
flow of dust.  In Section IV, we use the IRAS and 
Gaia triangular features to determine the velocity 
vectors on Earth of the four flows associated with 
the nearby caustic ring.  We also give rough estimates 
of their densities.  Section V provides a summary.

\section{The Caustic Ring Model}

The Caustic Ring Model is a proposal for the phase-space
distribution of the halo of the Milky Way, and the halos
of other disk galaxies.  The model is axially symmetric, 
and self-similar in its time evolution.  One of its 
distinguishing features is the presence of caustic rings 
of dark matter in the galactic plane.  In this section we 
give background information on dark matter caustics, a 
detailed description of caustic rings, the model predictions 
for the caustic ring radii, and the previously claimed 
observational evidence in support of the model. Finally 
we present additional evidence from the rotation curve 
of M31 and from a couple of triangular features in the 
Gaia skymap.

\subsection{Caustic rings}

\subsubsection{Dark matter caustics}

Caustics appear in a flow of energy or matter when two conditions
are satisfied: 1)  the flow is collisionless and 2) the velocity 
dispersion of the flow is small. Cold dark matter particles are 
collisionless - i.e. they have only gravitational interactions 
in excellent approximation - and have very small primordial 
velocity dispersion $\delta v$ \cite{sing,Natarajan}. Hence, 
caustics are expected in the distribution of dark matter.  They 
appear at the very moment multi-streaming begins.  Indeed, as 
mentioned already in Section I, cold dark matter particles lie 
on a thin three-dimensional hypersurface embedded in six-dimensional 
phase-space.   This phase-space sheet wraps around inhomogeneities 
such as galaxies. This results in a number of discrete flows through 
any point in physical space.  Caustics lie at the boundaries of 
regions with differing number of flows, one region having $K$ 
flows and the other $K+2$ flows.

Each particle in the phase-space sheet can be labeled by three 
parameters $(\alpha_1, \alpha_2, \alpha_3) \equiv \vec{\alpha}$. Let
$\vec{x}(\vec{\alpha},t)$ be the position of particle $\vec{\alpha}$ 
at time $t$. The number $K$ of discrete flows through a physical point 
$\vec{r}$ is the number of solutions $\alpha_j(\vec{r},t),~j=1,2, ..K$,
of the equation $\vec{r} = \vec{x}(\vec{\alpha},t)$.  In physical space, 
the number density is given by
\begin{equation}
n(\vec{r},t) = \sum_{j=1}^K
\frac{1}{|D(\vec{\alpha}_j(\vec{r},t),t)|}\;
\frac{d^3 N}{d\alpha^3}(\vec{\alpha}_j(\vec{r},t))  
\label{density}
\end{equation}
where $\frac{d^3 N}{d\alpha^3}(\vec{\alpha})$ is the particle 
number density in parameter space, and $D(\vec{\alpha},t) = 
\text{det}\Big(\frac{\partial\vec{x}}{\partial\vec{\alpha}}\Big)$ 
is the Jacobian of the transformation 
$\vec{\alpha} \rightarrow \vec{x}(\vec{\alpha},t)$.  
Eq.~(\ref{density}) shows that the physical space density 
diverges where the map $\vec{\alpha} \rightarrow \vec{x}(\vec{\alpha},t)$ is 
singular.

\subsubsection{Outer and inner caustics}

Cold dark matter particles fall in and out of the gravitational 
potential well of a galaxy numerous times over the galaxy's history.  
The in and out flows of the particles necessarily form inner and 
outer caustics \cite{sing,Natarajan}.  The outer caustics are 
topological spheres.  They appear near where the particles in 
an outflow are at their maximum distance from the galactic center 
before falling back in. The inner caustics appear near where the 
particles with the most angular momentum in an inflow are closest 
to the galactic center before falling back out. The structure of 
the inner caustics is determined by the velocity field of the 
particles at their last turnaround, i.e. when their radially 
outward flow was last stopped by the gravitational attraction 
of the galaxy. A shell containing such particles is called a 
turnaround sphere.  When the velocity field is irrotational 
($\vec{\nabla} \times \vec v =0$), the inner caustics have a
tent-like structure, described in ref. \cite{Natarajan}. When 
the velocity field is dominated by net overall rotation 
($\vec{\nabla} \times \vec v \neq 0$), the inner caustics 
are closed tubes with a \textit{tricusp} cross-section, 
called caustic rings.  The structure of caustics is stable 
under small perturbations in the velocity field \cite{Natarajan}.

The inner caustics provide a tool to differentiate between a 
rethermalizing Bose-Einstein condensate of axions, or ALPs, 
and the other dark matter candidates. Ordinary CDM, including 
WIMPs and sterile neutrinos, never acquires a velocity field 
with large scale rotation.  Indeed primordial rotational 
modes die out as a result of the Hubble expansion, setting 
$\vec{\nabla} \times \vec{v} = 0$ as an initial condition for all 
dark matter candidates.  Galactic halos acquire angular momentum 
through tidal torquing by nearby protogalaxies in the early phases 
of structure formation \cite{Peebles}.  Even though galactic halos 
acquire angular momentum, the velocity field of the dark matter 
particles remains irrotational if the dark matter is ordinary CDM 
\cite{Natarajan}.  The inner caustics have tent-like structure in 
that case.  Axions, as well as any axion-like particles (ALPs)
produced by the vacuum realignment mechanism \cite{axdm}, behave 
differently from ordinary CDM because they thermalize as a result 
of their gravitational self-interactions and form a Bose-Einstein 
condensate (BEC) \cite{CABEC,Erken,Christopherson}.  Dark matter axions 
rethermalize sufficiently fast by gravitational self-interactions while 
they fall in and out of a galactic gravitational potential well that 
almost all go to the lowest energy available state consistent with 
the angular momentum acquired by tidal torquing \cite{Erken,angmom}.  
That state is one of rigid rotation on the turnaround sphere.  As a
result dark matter axions fall in with a rotational velocity field 
and make caustic rings.  The observational evidence for caustic 
rings, discussed below, suggests therefore that the dark matter 
is axions, or ALPs, at least in part. If the dark matter is a 
mixture of axions and ordinary cold dark matter, caustic rings 
still form provided the fraction of axions is large enough.  
Ref. \cite{angmom} places a lower limit of $\sim$ 35\% on 
the axion fraction based on the prominence of features in 
galactic rotation curves associated with caustic rings.

\subsection{Caustic ring structure}

Here we briefly describe the properties of an axially and 
reflection symmetric caustic ring and the flows associated 
with it \cite{sing}. The axial coordinate being irrelevant, 
the dark matter particles are conveniently labeled by two 
parameters $(\alpha , \eta)$ where $\eta$ is the time when 
a given particle crosses the $z=0$ plane and $\alpha$ the 
angle from the $z=0$ plane at the time of the particle's 
most recent outer turnaround.  The coordinates of the 
particles near the caustic are given by 
\begin{eqnarray}
\rho &=& a + \frac{1}{2} u (\eta - \eta_{0})^{2} - 
\frac{1}{2} s \alpha^{2} 
\nonumber \\
z &=& b \alpha \eta 
\label{GAIA_rhoz} 
\end{eqnarray}
where $a, \; u, \; \eta_{0}, \; s$ and $b$ are constants 
characterizing the caustic ring.  Since actual caustic 
rings are only approximately axially symmetric, the 
five constants vary to some extent along the ring.
The caustic occurs where the Jacobian 
$|D_{2} (\alpha, \eta)| \equiv |\text{det} 
\frac{\partial(\rho, z)}{\partial(\alpha, \eta)}|$ 
is zero.  In the $\rho$-$z$ plane, it takes the shape 
of a tricusp, shown in Fig. 2.  The tricusp is the 
cross-section of a caustic ring.  Its sizes, $p$ in 
the $\rho$-direction and $q$ in the $z$-direction, 
are given by \cite{sing} 
\begin{eqnarray}
p &=& \frac{1}{2} u \eta_{0}^{2} \nonumber \\
q &=& \frac{\sqrt{27}}{4} \ \frac{1}{\sqrt{\zeta}} \ p \ , 
\label{GAIA_pq_defn}
\end{eqnarray}
where 
\begin{equation}
\zeta = \frac{su}{b^2} = \frac{27}{16} \ \frac{p^2}{q^2} \ . 
\label{GAIA_zeta_defn}
\end{equation}
If $\zeta = 1$, the tricusp is invariant under rotations 
by $120^\circ$ in the $\rho$-$z$ plane. 

It is convenient to write Eqs.~(\ref{GAIA_rhoz}) in terms 
of the rescaled coordinates 
$X = \frac{\rho - a}{p}$, $Z = \frac{z}{p}$:
\begin{eqnarray}
X &=& (T-1)^2 - A^2 \nonumber \\
Z &=& \frac{2}{\sqrt{\zeta}} \ A T 
\label{GAIA_XZAT}
\end{eqnarray}
where $A = \alpha \sqrt{\frac{s}{u \eta_0 ^2}}$ and 
$T = \frac{\eta}{\eta_0}$. For a given point $(X, Z)$ 
near the caustic, the parameters $(T,A)$ of particles
at that point are found by solving the quartic equation:
\begin{equation}
X = (T-1)^2 - \frac{\zeta Z^2}{4 T^2} \ , 
\label{GAIA_XT}
\end{equation}
and setting $A = {\sqrt{\zeta} Z \over 2 T}$.  Each real 
solution corresponds to a flow of dark matter particles 
through that point. There are two flows through each 
point outside the tricusp, and four flows through each 
point inside.  The density of the flow corresponding
to a real solution $(A_j,T_j)$, $j=1,..,K$ with 
$K$ = 2 or 4, is
\begin{equation}
d_j (\rho,z) = \frac{1}{2 p b \rho} \; \frac{dM}{d\Omega d\eta} \; 
{\cos(\alpha_j) \over |(T_j-1)T_j + A_j^2|} 
\label{density_formula}
\end{equation}
where $ \frac{dM}{d\Omega d\eta} $ is the infall rate, i.e. 
the mass of dark matter particles falling in per unit time 
per unit solid angle. The velocity vector of the $j$th flow 
has components
\begin{eqnarray}
v_{j\rho} &=& u \eta_0 (1-T_j) \nonumber \\
v_{j z} &=& - u \eta_0 \ \frac{Z}{2T_j}  \nonumber \\
v_{j \phi} &=& \sqrt{v^2 - v_{j\rho} ^2 - v_{j z} ^2} 
\label{GAIA_vrhozphi}
\end{eqnarray}
where $v$ is the speed of the flow, and the $\phi$-direction  
is $\hat{\phi} = \hat{\rho}\times\hat{z}$.  For an axially 
symmetric caustic ring whose center coincides with the 
Galactic Center, $\hat{\rho}$ points radially outward, 
$\hat{z}$ points toward the Galactic North Pole and 
$\hat{\phi}$ in the direction of Galactic rotation.

\subsection{Model properties}

The Caustic Ring Model is described in detail in ref. \cite{Duffy}.
Here we only list the properties directly relevant to our discussion 
of the Gaia triangles and the M31 rotation curve. The model assumes 
that the flow of dark matter particles in a galactic halo is 
self-similar in time \cite{FGB,STW}, and axially symmetric.  
It gives the overall phase-space distribution of a disk galaxy's 
halo in terms of the galaxy's rotation velocity $v_{\rm rot}$ and 
a parameter $j_{\rm max}$ which is a dimensionless measure of the 
galaxy's angular momentum.  The model predicts a disk galaxy to have 
caustic rings in the galactic plane at radii given approximately by 
\cite{crdm}:
\begin{equation}
a_n\simeq \frac{40\text{ kpc}}{n}
\left(\frac{v_{\text{rot}}}{220~\text{km/s}}\right)
\left(\frac{j_{\text{max}}}{0.18}\right) ~~\ ,
\label{crr}
\end{equation}
where $n = 1,2,3, ..$.  More precise predictions for the caustic
ring radii are given in ref. \cite{Duffy}.  The $n$th caustic 
ring forms in the flows of particles experiencing the $n$th 
inner turnaround in their history. The nominal values 
$v_{\rm rot} \simeq 220$ km/s and $j_{\rm max} \simeq 0.18$ 
apply to our Milky Way galaxy. The length scale in the 
$a_n \approx \frac{40\text{kpc}}{n}$ prediction for the 
Milky Way is set by the assumption that the radius of the 
solar orbit is 8.5 kpc.  Four rings are therefore predicted 
to lie outside the solar orbit and the fifth just inside.  The 
model is not so detailed as to predict the sizes $p$ and $q$ 
of the caustic ring cross-sections.  Values of $p/a$ estimated 
from rises in the Milky Way rotation curve range from 0.016 to 
0.1 \cite{MWcr}.

When the Caustic Ring Model was proposed, the observational evidence 
claimed for caustic rings raised a puzzle because caustic rings require
the velocity field to have net overall rotation whereas the velocity 
field of ordinary CDM is irrotational.  As mentioned already, axion 
Bose-Einstein condensation resolves this puzzle because cold dark matter 
axions thermalize sufficiently fast by gravitational self-interactions 
that almost all go to the lowest energy available state consistent with 
the angular momentum acquired by tidal torquing and that state is one of 
rigid rotation on the turnaround sphere.  It was shown in ref. \cite{case}
that axion Bose-Einstein condensation justifies the Caustic Ring Model 
in all its aspects, including the fact that the caustic rings lie 
in the galactic plane, the magnitude of the parameter $j_{\rm max}$ 
and the pattern, Eq.~(\ref{crr}), of caustic ring radii.

Ref. \cite{angmom} analyzed the behavior of the vortices that form 
in the axion BEC constituting galactic halos.  Unlike the vortices in 
superfluid $^4$He, the vortices in axion BEC are mutually attractive
and combine into one Big Vortex along the rotation axis of a disk 
galaxy.  This results in a modification of the Caustic Ring Model 
since the infall was assumed to be isotropic \cite{Duffy} in the 
original formulation. The presence of a Big Vortex causes a 
depletion along the rotation axis and a compensating enhancement 
along the equatorial plane.  The enhancement along the equatorial 
plane serves to explain why the bumps in galactic rotation curves 
attributed to caustic rings are so prominent.  It had been noted 
in ref. \cite{MWcr} that the rises in the Milky Way rotation curve 
attributed to caustic rings are approximately a factor 4 larger 
than expected.  The presence of a Big Vortex resolves this puzzle 
because caustic rings are formed in the flows of particles falling 
in and out close to the Galactic Plane where the density is enhanced.  
When estimating the densities of the flows associated with the nearby 
5th caustic ring in Section IV, we will assume that there is a Big 
Vortex and that the densities are enhanced by a factor 4 compared
to the predictions of ref. \cite{Duffy}. 

\subsection{Summary of previous evidence}

Several observations have been previously claimed as supporting
evidence for the Caustic Ring Model.  We briefly review these
observations here.

\subsubsection{Combined rotation curve}

Caustic rings produce bumps at their locations in the rotation 
curves of galaxies \cite{crdm,sing}. The extended and well measured 
rotation curves of 32 galaxies, published in refs.~\cite{BeSa}, were 
analyzed in ref.~\cite{Kinney}. The radial coordinate $r$ in 
each rotation curve was rescaled according to 
\begin{equation}
r \rightarrow \tilde{r}
= r \left(\frac{220\text{ km/s}}{v_{\text{rot}}}\right) \; ,
\label{resc}
\end{equation}
to remove the dependence of the caustic ring radii on 
$v_{\rm rot}$; see Eq.~(\ref{crr}).  The rescaled rotation 
curves were then added to each other. The combined rotation 
curve, shown in Fig. 3, has two peaks, one near $40$ kpc 
and one near $20$ kpc, with significance of $3.0 \sigma$ 
and $2.6\sigma$ respectively. The presence and locations 
of the two peaks are explained if they are caused by the 
$n=1$ and $n=2$ caustic rings of dark matter in those 
galaxies.

\subsubsection{Milky Way rotation curve \cite{MWcr}}

The inner ($r<r_\odot$) Milky Way rotation curve derived 
\cite{Clemens} from the Massachusetts-Stony Brook North 
Galactic Plane CO Survey \cite{CO} shows a series of sharp 
rises between $3$ and $8.5$ kpc.  Each rise starts with an 
upward kink and ends with a downward kink as expected for 
rises caused by caustic rings of dark matter \cite{sing}.  
The Caustic Ring Model predicts ten rises between 3 and 
8.5 kpc, assuming the value of $j_{\rm max} = 0.18$ 
derived earlier from the ratio of baryonic to dark 
matter contributions to the Milky Way rotation curve 
at the solar radius \cite{STW}\cite{Duffy}.  Allowing 
for ambiguities in identifying rises, the number of rises 
in the rotation curve between 3 and 8.5 kpc is in fact 
ten plus or minus one.  When the predicted caustic ring 
radii are fitted to the radii where the rises start in the 
rotation curve, with $j_{\rm max}$ the only adjustable 
parameter, the remaining root mean square relative 
discrepancy is 3.1\% \cite{MWcr}.

The outer rotation curve ($r>r_{\odot}$) is less well 
measured, but it does have a prominent rise between 12.7 
and 13.7 kpc \cite{Olling,Binney}, which is the predicted 
location of the third caustic ring. Furthermore a ring of 
stars, named the ``Monoceros Ring", was discovered in the 
Galactic Plane at $r \sim 20$ kpc \cite{Newberg}. It is 
shown in refs.~\cite{Monoceros}\cite{starsgas} that this 
ring of stars is a plausible outcome of the second caustic 
ring's gravitational field. 

\subsubsection{IRAS triangle \cite{MWcr}}

Looking in the direction tangent to a caustic ring, one may 
expect to observe the imprint of the ring's tricusp cross 
section on ordinary matter. A triangular feature seen in 
the IRAS map in the direction of Galactic Coordinates 
$(l,b)=(80^\circ,0^\circ)$ has been interpreted as the 
imprint of the $5$th caustic ring. Relevant parts of 
the IRAS map can be found at 
\url{https://www.phys.ufl.edu/~sikivie/triangle/}. The 
orientation of the IRAS triangle with respect to the 
Galactic Center and the Galactic Plane is consistent 
with the Caustic Ring Model.  Also, the position of 
the triangle on the sky matches the position of the 
sharp rise (between 8.28 and 8.43 kpc) in the Milky Way 
rotation curve attributed to the $5$th caustic ring. 
However, no triangular feature along the other tangent 
direction to the $5$th caustic ring was found in the 
IRAS map.

\subsection{M31 rotation curve}

Our nearest major Galactic Neighbor M31 is viewed almost edge-on 
from our location.  Fig. 4 shows the rotation curve of M31
derived by Chemin et al. \cite{Chemin} from their HI survey of that 
galaxy.  There are several bumps in this rotation curve on both the 
receding and approaching sides. The first or outermost bumps are 
approximately at $30$ kpc, the second ones at $15$ kpc, 
and the third at $10$ kpc. These numbers are consistent with 
the relation $a_n \sim \frac{1}{n}$ predicted by the Caustic 
Ring Model. The first bumps near $30$ kpc are well outside the 
last observed spiral arm (at $r \sim 25$ kpc) and they match 
perfectly on both sides.  This strongly indicates the presence 
of a ring-like structure in the dark halo. We also note that, 
although the first bumps match perfectly, the second ones appear 
at slightly different radii on one side than the other.  And the 
third bump from the receding side is missing. These discrepancies 
give an indication of how closely we may expect the idealizations 
of the model to be matched by reality.  The fact that the third 
caustic ring does not show on the receding side may mean that 
it lies somewhat outside of the galactic plane there.  It is 
worth mentioning that there are several sources for the rotation 
curve of M31 at smaller radii, summarized and compared in 
Ref.~\cite{Sofue}, and they do not all agree with each other.

\subsection{Gaia triangles}

We mentioned that the triangular feature seen in the 
IRAS map in the direction $(l,b)=(80^\circ,0^\circ)$ 
is not matched by a similar feature on the other side, 
i.e. at $(-80^{\circ},0^{\circ})$.  However, the sky 
map recently released by the Gaia Collaboration
\cite{Gaia1,Gaia2} has a triangular feature at 
$(l,b) = (-91^{\circ},0^{\circ})$ as well as a 
feature at $(80^{\circ},0^{\circ})$ matching the 
triangular feature in the IRAS map. See:
\url{https://www.phys.ufl.edu/~sikivie/Gaiamap/} .
We refer to the features as right and left triangles
respectively.  Relevant parts of the Gaia skymap 
are reproduced in Figs. 5 and 6.  The shape and the 
orientation of the right triangle are consistent 
with the expected shape and orientation of a caustic 
ring imprint.  The right triangle is farther from the 
Galactic Center and approximately 40\% smaller than 
the triangle on the left, but we believe both are 
imprints of the $5$th caustic ring. We attribute 
the asymmetry in location to a displacement, by 
$5.8^\circ$ to the right, of the center of the 
caustic ring relative to the Galactic Center; 
see Fig. 7 for an illustration.  The possibility
that the ring is not a perfect circle between the 
two tangent points is discussed in Section IV.B.

The triangular features in the Gaia map are darker 
than their surroundings.  Obscuration is generally 
due to dust.  The obscuration is more uniform over 
the Gaia right triangle than over the Gaia left triangle.  
Assuming the background of star light to be uniform, we 
estimate the optical depth for light absorption $\tau_d$ 
is of order 1.7 over the area of the right triangle, 
and varying approximately from 0.9 to 3.0 over the area 
of the left triangle.  The 3D dust map constructed in 
ref.~\cite{Green1} on the basis on Gaia, PANSTAR-1 
and 2MASS data, also shows a feature in the direction 
of the left triangle; see Fig. 8. Unfortunately, 
the direction of the right triangle is not covered by 
this map. For a better understanding of the distribution 
of dust in the direction of the left triangle, 
Fig. 9 shows the accumulated reddening in 
successive distance segments. Panels (b) and 
(c) of this figure have the most accumulated dust in 
the direction of interest.  It seems that most of the 
dust making up the feature in Fig. 8 is situated 
between $0.8$ kpc and $1.6$ kpc. This is consistent with 
the location of the caustic ring in the left tangent 
direction based on the calculations in Section IV.

We performed an analysis of the statistical significance 
of the left and right triangular features in the Gaia 
skymap.  The analysis is described in Appendix A.  

When the IRAS triangle was first discussed as a
candidate imprint of the 5th caustic ring \cite{MWcr},
it was proposed that this feature is produced by gas and 
dust in thermal equilibrium in the gravitational field 
of the 5th caustic ring.  However in ref. \cite{starsgas} 
it is shown that the features produced in this way are 
not as sharp as the observed features. Here we propose 
instead that the features are produced by dust that is 
entrained by the axion flows forming the caustic ring.  
The entrained dust particles have the same trajectories 
as the axions and thus make the same caustics.  This 
would explain why the triangular features are sharp. 
Section III discusses the entrainment of dust particles
by cold axion flows.

We also propose here an explanation why the right 
triangle does not show in the IRAS map whereas the 
left triangle does show.  The caustic ring in the 
left tangent direction lies in the midst of the 
stellar activity associated with the Orion Spur 
of the Sagittarius Spiral Arm.  The dust is heated 
there by stars and reradiates the heat in the 
infrared, where IRAS is sensitive.  The caustic 
ring in the right tangent direction lies in a 
relatively quiet place.  Its dust does not get 
heated sufficiently to show up in the IRAS map.  

\section{Dust entrainment}

In this section we discuss physical processes that 
may be responsible for the formation of the triangles 
observed in the IRAS and Gaia maps of the Milky Way.  As 
mentioned in the previous section, we interpret the 
triangles as due to the entrainment of dust by the 
axion flows forming the nearby caustic ring.  The dust
particles follow the same trajectories as the axions 
and hence form the same caustics.  We propose this as 
an explanation for the sharpness of the triangles. 
For the explanation to be plausible, it must be shown 
not only that dust is entrained by cold axion flows but 
also that the entrainment occurs in spite of the drag 
on the dust due to gas in the Galactic Disk.  We first 
derive a formula for the friction on a dust particle 
moving with respect to a highly degenerate Bose-Einstein 
condensed axion fluid.  The frictional force is inversely 
proportional to the temperature $T$ of the axions. We next 
discuss the temperature that axion dark matter has today, 
and compare the friction of the axion fluid with the drag 
due to gas.   Finally we consider dust-dust collisions and 
find that they plausibly explain why the features seen in 
the IRAS and Gaia maps are triangular rather than tricuspy.

\subsection{Friction on a cold axion fluid}

Cold dark matter axions thermalize by gravitational
self-interactions in a regime, called the ``condensed
regime", where their energy dispersion is much less than 
their thermalization rate.  Their thermalization rate
is of order \cite{CABEC,Erken}
\begin{equation}
\Gamma \sim 4 \pi G n m^2 \ell^2
\label{therm}
\end{equation}
where $G$ is Newton's gravitational constant, $m$ is 
the axion mass, $n$ is the number density of axions, 
and $\ell = {1 \over \delta p}$ their correlation length.  
$\delta p$ is the momentum dispersion of the axion fluid. 
By definition, $\Gamma = 1/\tau$ where $\tau$ is the time 
scale over which each axion may change its momentum by
order $\delta p$. 

\subsubsection{Scale of inhomogeneity vs. correlation length}

The correlation length $\ell$ is unrelated to the scale 
of homogeneity and should not be confused with it \cite{SC}.
To make precise this distinction consider a generic state 
of the cold axion dark matter fluid.  Almost all axions are 
in a small number of particle states with almost identical 
wavefunctions.  Let $\Psi_0(\vec{x}, t)$ be the wavefunction 
of one of those highly occupied states.  It defines a flow of 
density $n(\vec{x}, t) = N_0 |\Psi_0(\vec{x}, t)|^2$, where 
$N_0$ is the number of particles in the state, and velocity
field $\vec{v}(\vec{x}, t) = {1 \over m}
\vec{\nabla} Im \ln \Psi_0(\vec{x}, t)$.  Starting with an 
arbitrary $\Psi_0$, one may construct a complete orthonormal 
set of wavefunctions
\cite{SC}
\begin{equation}
\Psi_{\vec{k}}(\vec{x}, t) = \Psi_0(\vec{x}, t)
e^{i \vec{k} \cdot \vec{\chi}(\vec{x}, t)}
\label{ONC}
\end{equation}
where the $\vec{\chi}(\vec{x}, t)$ are the co-moving coordinates 
implied by the velocity field $\vec{v}(\vec{x}, t)$ and the
requirement that the particle density is constant in co-moving
coordinate space.  For small $\vec{k}$ the particle states   
$\Psi_{\vec{k}}$ are very close to $\Psi_0$, having the same
density field and almost the same velocity field.  We may write
the axion field in the non-relativistic limit as
\begin{equation}
\phi(\vec{x}, t) = \sum_{\vec{k}} {1 \over \sqrt{2m}}
[e^{- i m t} \Psi_{\vec{k}}(\vec{x}, t) a_{\vec{k}}(t) 
+ e^{ i m t} \Psi_{\vec{k}}^*(\vec{x}, t)
a_{\vec{k}}^\dagger(t)]
\label{expand}
\end{equation}
where $a_{\vec{k}}(t)$ and $a_{\vec{k}}^\dagger(t)$ are
annihilation and creation operators satisfying canonical
equal-time commutation relations.  Let  $N_{\vec{k}} =
\langle \Phi| a_{\vec{k}}^\dagger a_{\vec{k}}|\Phi \rangle$ be the
particle state occupation numbers in a state $|\Phi\rangle$ of the
axion fluid.  The correlation length in comoving coordinates is
$\ell = {1 \over \delta k}$ where
\begin{equation}
\delta k = \sqrt{{1 \over N} \sum_{\vec{k}}
\vec{k}\cdot\vec{k} N_{\vec{k}}}
\label{momdis}
\end{equation}
and $N = \sum_{\vec{k}} N_{\vec{k}}$ is the total number  
of particles.  If a Bose-Einstein condensate (BEC) forms, 
a fraction of order one of the total number $N$ of particles 
go to the same particle state, e.g. the state of wavefunction 
$\Psi_0$.  In that case the two-point axion field equal-time 
correlation function is \cite{SC}
\begin{equation}
\langle \Phi| \phi(\vec{x}, t) \phi(\vec{y}, t) | \Phi \rangle
= {N_0 \over 2 m}(\Psi_0^*(\vec{x}, t) \Psi_0(\vec{y}, t) + c.c.)
+ ...
\label{corrf}   
\end{equation}
where $N_0 \sim N$ is the number of particles in the condensate, and
the dots represent the contributions, which fall off exponentially
or as a power law, of particles that are not in the condensate.
Eq. (\ref{corrf}) shows that a BEC is `perfectly' correlated over
its full extent, i.e. its correlation length $\ell$ is the size
of the region over which $\Psi_0(\vec{x}, t)$ has support.  In
contrast, the scale of inhomogeneity $s$ of the condensate is
the distance scale over which the density $n(\vec{r}, t)$ varies
by order one.  $\ell$ can be arbitrarily large compared to $s$.
In the axion case, $s$ may be the size of mini-clusters or of
galaxies whereas $\ell$ may be the size of the horizon.

\subsubsection{Thermal relaxation}

In estimating the frictional force of the axion fluid on a dust 
particle, we will assume that the axion fluid has thermalized 
completely.  However, it is not clear to us to what extent 
the axion fluid has thermalized.  The assumption of complete 
thermalization is made to allow us to make an estimate of the 
aforementioned drag.  If fully thermalized the axion fluid 
consists of a BEC of $N_0$ axions in a single state, of 
wavefunction $\Psi_0$, plus a thermal distribution of axions 
with chemical potential $\mu = m$ and temperature $T$. The 
wavefunction $\Psi_0$ is time-dependent on the Hubble time 
scale since this is the time scale over which large scale 
structure grows by gravitational instability.  The thermal 
relaxation time scale is much shorter than that as we now 
show in the case, used as an example, of the cold axion flow 
forming the nearby caustic ring.  The flow extends over a region 
of order 100 kpc in size.  Its correlation length is that size, 
$\ell \sim$ 100 kpc, regardless of the inhomogeneities in it.  
Its average energy density $nm \sim 10^{-26}$ gr/cc over that 
length scale \cite{Duffy}. Eq.~(\ref{therm}) implies that its 
relaxation rate is 
\begin{equation}
\Gamma \sim {10^4 \over {\rm sec}} 
\left({m \over 10^{-5}~{\rm eV}}\right)~~\ .
\label{rate}
\end{equation}
The relaxation rate exceeds the Hubble rate by some 21 orders 
of magnitude.  

The relaxation process can be described heuristically as follows.
The axion fluid consists of a huge number of axions occupying a small 
number of states with the wavefunctions given in Eq.~(\ref{ONC}).  The 
average gravitational field produced by the axion fluid is 
$g(\vec{x},t)$ sourced by the density $m n(\vec{x},t) = 
m N_0 |\Psi_0(\vec{x},t)|^2$. However, the actual gravitational 
field fluctuates around this average because the actual density 
is modulated over a distance scale $\ell = 1/\delta k$.  The 
root mean square deviation of the gravitational field from 
its average value $g(\vec{x},t)$ is 
\begin{equation}
\sigma_g \sim 4 \pi G n m \ell~~\ .
\label{grav}
\end{equation}
As discussed in ref. \cite{Erken}, $\sigma_g$ is the outcome of 
a random walk, the sum of many terms with random phases.  The 
average gravitational field $g(\vec{x},t)$ determines the dynamical 
evolution of the axion fluid.  Since we are interested in the 
relaxation of the axion fluid, as opposed to its overall dynamical 
evolution, we ignore $g(\vec{x},t)$ henceforth.  We set it equal 
to zero by adopting a reference frame in which the fluid is freely 
falling. 

The gravitational field fluctuations change all the particle momenta by 
amounts of order $\delta p$ in a time $\tau \sim {\delta p \over m \sigma_g}$. 
Substituting Eq.~(\ref{grav}) yields Eq.~(\ref{therm}) as an estimate
of the relaxation rate.  Relaxation results in momentum distributions 
of ever increasing likelihood.  Although far from thermal equilibrium 
to start with, the axion fluid may reach near-thermal equilibrium by 
gravitational self-interactions before the present.  It may also 
absorb heat from other species, e.g. from baryons \cite{Erken}.  
There is however a maximum temperature the axion fluid may become 
fully thermalized at by gravitational self-interactions.  Since 
the axions may change their momenta by order $\delta p$ in a time 
of order $\tau$, they can at most reach velocities of order 
\begin{equation} 
v_m \sim \sigma_g t_0 \sim 4 \pi G n m t_0 \ell 
\sim 0.4~\left({\Omega_a \over 0.27}\right)
\left({\ell \over t_0}\right)~~\ .
\label{vm}
\end{equation}
Here we have set $nm = \Omega_a \rho_{\rm crit}(t_0)$, where
$\rho_{\rm crit}(t_0)$ is the present critical energy density 
for closing the universe and $\Omega_a$ is the fraction thereof 
in axions.  This implies a maximum temperature 
\begin{equation}
T_m \sim {1 \over 3} m v_m^2 \sim 
6 \cdot 10^{-2}~m \left({\ell \over t_0}\right)^2 
\left({\Omega_a \over 0.27}\right)^2 
\label{Tm}
\end{equation}
that axions may become fully thermalized at before today.

\subsubsection{Frictional force}

Since the gravitational field produced by the axion fluid 
at any space-time point is a sum of many terms with random 
phases, with $\sigma_g$ the average outcome, the probability 
that the gravitational field has value between $g$ and 
$g + dg$ is 
\begin{equation}
{\cal P}(g) dg = {1 \over \sqrt{2 \pi} \sigma_g}
e^{-{1 \over 2}({g \over \sigma_g})^2}~dg
\label{prob}
\end{equation}
by the central limit theorem.  Eq.~(\ref{prob}) holds in the 
absence of the dust particle, whose presence we have ignored 
so far. In the presence of the dust particle, the random walk 
becomes biased toward gravitational fields that slow down the 
dust particle with respect to the axion fluid because such 
slowdown is accompanied by an increase in the energy, and 
therefore the entropy, of the axion fluid.  Let $M$ be the
mass of the dust particle and $\vec{v} = v \hat{n}$ its
velocity with respect to the axion fluid.  A gravitational 
field of strength $-g \hat{n}$ for the duration $\tau$ 
slows down the dust particle by $\Delta v = - g \tau$ 
and therefore increases the energy of the axion 
fluid by $\Delta E = M v |\Delta v| = M g \tau v$. 
Its entropy increases by $\Delta S = {1 \over T} \Delta E$.  
As Boltzmann pointed out, a $\Delta S$ increase in entropy 
signifies an increase by the factor $e^{\Delta S}$ in the 
number of available microstates and therefore an increase 
by that factor in the relative probability to have the 
gravitational field that causes it.  Thus in the presence 
of the dust particle, the probability distribution of the 
gravitational field at the dust particle's location is 
modified from Eq.~(\ref{prob}) to
\begin{equation}
{\cal P}^\prime(g) dg = 
C e^{-{1 \over 2}({g \over \sigma_g})^2 + \Delta S}~dg
\label{mprob}
\end{equation}
where
\begin{equation}
\Delta S = {1 \over T} M g \tau v = 
{M v \over m T \ell}{g \over \sigma_g}
\label{DS}
\end{equation}
and $C$ is a normalization constant determined by the 
requirement that the total probability is one.  The 
deceleration $d$ of the dust particle is the average 
of the $g$-distribution in Eq.~(\ref{mprob}).  One
readily finds
\begin{equation}
d = {M v \over m T \ell} \sigma_g
\sim 4 \pi G n M v {1 \over T}~~\ .
\label{dec}
\end{equation}
This formula for friction is different, and applies 
in different circumstances, from the standard formula 
for dynamical friction \cite{Chandra,BT}. Both formulae, 
Eq.~(\ref{dec}) and the standard formula whose RHS is 
$4 \pi G^2 n m M v^{-2} \ln(\Lambda)$ where $\Lambda$ 
is the ratio of maximum to minimum impact parameters, 
describe the drag on a heavy mass $M$ moving through 
a fluid, as a result of the gravitational interactions of 
the heavy mass with the particles in the fluid.  The standard 
formula assumes that the particles in the fluid do not interact 
among themselves. Only their gravitational interaction with the 
heavy mass is taken into account.  They get scattered by the 
heavy mass plowing through and as a result remove kinetic energy 
from it, slowing it down.  Eq.~(\ref{dec}) assumes instead 
that the particles in the fluid interact with one another 
sufficiently strongly that they thermalize while interacting 
gravitationally with the mass $M$.  

Eq.~(\ref{dec}) is not expected to be valid when $M$ is 
large, such as the mass of a star or even a small planet, 
because the Gaussian distribution in Eq.~(\ref{mprob}) 
does not extend to arbitrarily large values of $g$. For 
example, for $M = 10^{25}$ gr, $m = 10^{-5}$ eV, $T = 10^{-9}$ eV,
and $\ell = 100$ kpc, Eq.~(\ref{dec}) would need the Gaussian to 
extend to $g \sim 10^{44} v \sigma_g$, which it cannot possibly 
reach by the aforementioned random walk.

\subsection{Axion temperature}

The correlation length $\ell$ of the cold axion fluid, just 
after it was produced by vacuum realignment during the QCD 
phase transition, is of order the horizon at that time.  
Subsequently $\ell$ is stretched by the expansion of the 
universe.  The relaxation rate $\Gamma$, decreasing as 
$a(t)^{-1}$ where $a(t)$ is the scale factor, exceeds the 
Hubble expansion rate $H(t) = {1 \over 2 t}$ when the photon 
temperature is of order 1 keV \cite{CABEC,Erken}.  The 
cold axions form a BEC then and $\ell$ grows to be of 
order the horizon at that time.  Whereas Bose-Einstein 
condensation occurs on the $\tau$ time scale, full 
thermalization takes much longer \cite{Erken,BJ}.
Nonetheless, as was indicated above, nearly full 
thermalization may occur before the present and 
temperatures as high as $6 \cdot 10^{-2}~m$ may 
possibly be reached. The final temperature depends 
on the amount of heat that the axion fluid absorbs
and thermalizes.

\subsubsection{Heat from axions} 

The first and most obvious source of heat is the kinetic 
energy the cold axions themselves have because the axion 
field is inhomogeneous on the horizon scale at the QCD 
phase transition.   We assume here that the axion field 
was not homogenized by inflation,  so that the kinetic 
energy of cold axions is the highest possible.  We will 
see that even then the heat associated with the initial
kinetic energy of cold axions is negligible compared to 
the heat the axion fluid absorbs by cooling baryons (see 
below.).  The axion kinetic energy density is  
\begin{equation}
\rho_{a,{\rm kin}}(t) = {(\delta p (t))^2 \over 2 m} n(t) 
\sim \Omega_a \rho_{\rm crit}(t_0) {1 \over 2 t_1^2 m^2}
{a(t_1)^2 \over a(t)^5}
\label{akin}
\end{equation}
where $a(t)$ is the cosmological scale 
factor normalized such that $a(t_0) = 1$, and \cite{axdm}
\begin{equation}
t_1 \simeq 1.7 \cdot 10^{-7}~{\rm sec} 
\left({10^{-5}~{\rm eV} \over m}\right)^{1 \over 3}
\label{t1}
\end{equation}
is the time at which the axion mass effectively turns on 
during the QCD phase transition.  The last statement in 
Eq.~(\ref{akin}) follows from 
$\delta p(t) \sim {1 \over t_1} {a(t_1) \over a(t)}$ as 
is the case if inflation does not homogenize the axion 
field.  Let us call $t_*$ the time when cold axions 
fully thermalize among themselves.  They have at that 
time temperature $T_*$:
\begin{equation}
\rho_{a,{\rm kin}}(t_*) = 0.128 (m T_*)^{3 \over 2} T_*
\label{encon}
\end{equation}
provided $T_* << m$, as will be the case.  Combining 
Eqs.~(\ref{akin}) and (\ref{encon}) yields 
\begin{equation}
T = T_* a(t_*)^2 \sim
10^{-14}~{\rm eV}~\Omega_a^{2 \over 5} 
\left({10^{-5}~{\rm eV} \over m}\right)^{19 \over 15}
\label{aT}
\end{equation}
for the axion temperature today.   Note that $T$ does not 
depend on $t_*$ because the axions are non-relativistic
both before and after they thermalize.

\subsubsection{Heat from baryons}

The cold axion fluid may absorb heat from other species.
Ref.~\cite{Li} considered the possibility that the axions 
cool the photons and offered this as an explanation for the 
Li anomaly in primordial nucleosynthesis.  However, photon 
cooling can only occur marginally because it requires the 
correlation length to be as large as the horizon whereas 
by causality the correlation length must be at least 
somewhat shorter.  The observations of the cosmic microwave
background anisotropies by the Planck Collaboration indicate 
an effective number of neutrinos \cite{Planck} consistent 
with the absence of photon cooling.  So we ignore heat from 
photon cooling.  On the other hand, cooling of baryons by 
axion BEC occurs robustly according to the arguments of 
ref. \cite{Erken}.  The reported observation by the EDGES 
Collaboration \cite{EDGES} of the trough in the cosmic 
microwave radiation spectrum due to its absorption by 
neutral hydrogen at cosmic dawn indicates that baryons 
are significantly colder at that time than expected under 
standard assumptions.  The EDGES observation may be viewed 
as confirmation that axion BEC did indeed cool baryons 
\cite{Tmat,Houston}.

If axions and baryons reach full kinetic equilibrium before
today, a lower limit on the heat transfer from baryons to 
axions is the kinetic energy that baryons would have today
in the absence of axion cooling.  Keeping in mind that 
baryons and photons decouple from each other at a redshift 
$z_{\rm dec} \sim$ 160, the baryon kinetic energy density 
today, in the absence of axion cooling, is
\begin{equation}
\rho_{b,{\rm kin}}(t_0) \simeq {3 \over 2} T_\gamma(t_0)
\Omega_b \rho_{\rm crit}(t_0) {1 \over m_b} {1 \over 1 + z_{\rm dec}}~~\ ,
\label{bkin}
\end{equation}
where $T_\gamma(t_0) \simeq 2.7$ K is the present photon 
temperature, $\Omega_b \simeq 0.05$ is the present energy 
density fraction in baryons, and $m_b \simeq$ GeV is an 
average baryon mass.  Using Eq.~(\ref{encon}), baryon 
cooling by axions implies 
\begin{equation}
T > 0.7 \cdot 10^{-7}~{\rm eV} 
\left({10^{-5}~{\rm eV} \over m}\right)^{3 \over 5}
~~\ ,
\label{Tb}
\end{equation}
if axions and baryons reach full kinetic equilibrium.
The RHS of Eq.~(\ref{Tb}) exceeds, or saturates depending 
on the axion mass, our earlier estimate Eq.~(\ref{Tm}) 
of the highest temperature that the axion fluid may 
become fully thermalized at. 

\subsection{Dust entrainment}

As mentioned in Section II, dark matter axions rethermalize 
sufficiently fast by gravitational self-interactions while 
they fall in and out of a galactic gravitational potential 
well that they almost all go to the lowest energy state 
consistent with the angular momentum they have acquired 
by tidal torquing interactions with neighboring galaxies. 
That state is one of rigid rotation on the turnaround 
sphere.  As a result dark matter axions fall in with a 
rotational velocity field and make caustic rings.  

We will assume here, for simplicity, that the dark matter 
is entirely axions, or axion-like particles (as opposed
to a mixture of axions and ordinary CDM).  The infalling 
axions entrain gas and dust, but not stars for the reason 
stated at the end of subsection III.A.3.  The gas is not 
collisionless, of course.  Gas falling in with the axions 
collides with gas already in the galaxy and soon leaves 
the phase-space sheet on which the axions lie.  Whereas 
the  angular momentum of the gas is approximately conserved, 
the kinetic energy associated with its radial motion is 
dissipated into radiation.  As a result the gas settles 
in a rotating disk, where it participates in star formation.  
The stars formed rotate along with the gas.  Dust is  
produced in the late stages of stellar evolution.

We consider dust particles of typical size 
$D \sim 5 \cdot 10^{-5}$ cm \cite{Draine}, and mass 
$M \sim 3 \cdot 10^{-13}$ gr. Eq.~(\ref{dec}) implies 
that the speed of a dust particle relative to the axion 
flow decreases according to $v(t) = v(0) e^{- \gamma t}$ 
with  
\begin{equation}
\gamma \sim 4 \pi G \rho {M \over m}{1 \over T} 
\sim {4 \cdot 10^4 \over t_0} 
\left({\rho \over 10^{-26}~{\rm gr/cc}}\right)
\left({10^{-5}~{\rm eV} \over m}\right)
\left({M \over 3 \cdot 10^{-13}~{\rm gr}}\right)
\left({10^{-9}~{\rm eV} \over T}\right)~~\ , 
\end{equation}
implying that the dust particle is entrained in the 
absence of any other forces acting on it, even if the 
axion temperature is as high as we believe it can be.  
We now consider the effect of dust-gas and dust-dust
collisions.

\subsubsection{Dust-gas collisions}

The density of gas in the solar neighborhood is approximately 
$\rho_g \sim 3 \cdot 10^{-24}$ gr/cc, comprising of order one 
atom or molecule per cm$^3$ \cite{BT}.  The cross-section for 
hard-scattering on a dust particle is of order 
$\sigma \sim D^2 \sim 2.5 \cdot 10^{-9}$ cm$^2$.  A dust 
particle moving with velocity $v_g$ relative to the gas 
experiences a deceleration
\begin{eqnarray}
d_g &\sim& {\rho_g \sigma (v_g)^2 \over M}
= 2.2 \cdot 10^{-5}~{{\rm cm} \over {\rm sec}^2}
\left({\rho_g \over 3 \cdot 10^{-24}~{\rm gr/cc}}\right)\cdot
\nonumber\\ &\cdot&
\left({\sigma \over 2.5 \cdot 10^{-9}~{\rm cm}^2}\right)
\left({3 \cdot 10^{-13}~{\rm gr} \over M}\right)
\left({v_g \over 300~{\rm km/s}}\right)^2~~\ .
\label{dg}
\end{eqnarray}
At the caustic the axions flow relative to the gas with 
velocity 300 km/s in the direction of Galactic Rotation.
The speed of the dust particle with respect to the axion 
flow is $v = 300~{\rm km/s} - v_g$.  The acceleration of 
the dust particle in the direction of Galactic Rotation 
due to its friction on the axion flow is 
\begin{equation}
d \sim 2.7 \cdot 10^{-6}~{{\rm cm} \over {\rm sec}^2}
\left({\rho \over 10^{-26}~{\rm gr/cc}}\right)
\left({10^{-5}~{\rm eV} \over m}\right)
\left({M \over 3 \cdot 10^{-13}~{\rm gr}}\right)
\left({10^{-9}~{\rm eV} \over T}\right)
\left({v \over 300~{\rm km/s}}\right)
\label{da}
\end{equation}
according to Eq.~(\ref{dec}).  Setting $d_g = d$
yields a second order polynomial equation for $v_g$
whose relevant solution is
\begin{equation}
v_g = [\sqrt{\xi^2 + 2 \xi} - \xi]~300~{\rm km/s}
\label{vg}
\end{equation}
with 
\begin{eqnarray}
\xi &\sim& 6 \cdot 10^{-2}   
\left({\rho \over 10^{-26}~{\rm gr/cc}}\right)
\left({10^{-5}~{\rm eV} \over m}\right)
\left({M \over 3 \cdot 10^{-13}~{\rm gr}}\right)^2\cdot
\nonumber\\
&\cdot&
\left({10^{-9}~{\rm eV} \over T}\right)
\left({2.5 \cdot 10^{-9}~{\rm cm}^2 \over \sigma}\right)
\left({3 \cdot 10^{-24}~{\rm gr/cc} \over \rho_g}\right)~~\ .
\label{xi}
\end{eqnarray}
Several factors on the RHS of Eq.~(\ref{xi}) are poorly
known, including the axion mass and the temperature
of the axion fluid.  We consider two specific cases 
for illustrative purposes.  If $m \sim 10^{-5}$ eV 
and the axion fluid is not heated by the baryons, 
or anything else, so that $T \sim 10^{-14}$ eV, 
$\xi \sim 6 \cdot 10^3$ and therefore $v \sim $ 26 m/s.
The dust particle follows the axion fluid very closely
in this case.  If on the other hand $m \sim 10^{-6}$ eV 
and $T \sim 6 \cdot 10^{-8}$ eV, the latter being our 
estimate Eq.~(\ref{Tm}) of the highest temperature the 
axions may become thermalized at when $m \sim 10^{-6}$ 
eV, then $\xi \sim 0.01$ and $v_g \sim 39$ km/s.  
The dust particles move more slowly than the axion 
fluid in this case;  their velocity is 259 km/s in 
the Galactic Rest Frame vs. 520 km/s for the axion 
fluid. However, even though they move more slowly, 
we may still expect the dust particles to follow the 
same trajectories as the axion fluid and hence form 
the same caustics.

Collisions with gas diffuses the dust flow by imparting 
random transverse velocities to the dust particles, but 
not so much as to prevent caustic formation.  Indeed, 
the number of dust-gas collisions during one pass of 
a dust particle through the Galactic Disk is order 
$10^{14}$, each collision producing a random transverse 
velocity of order $v_g m_g/M$, which is less than 
$3 \cdot 10^{-10}$ km/s.  So the rms transverse 
velocity acquired is less than 3 m/s.

\subsubsection{Dust-dust collisions}

The density $n_d$ of the dust flows forming the triangular 
features in the Gaia map may be estimated from the optical 
depth $\tau_d \sim 2$ for the absorption of light over the 
area of the triangles.  The absorption length $\lambda \sim 
{1 \over n_d \sigma} \sim L/\tau_d$ where $L \sim$ 1 kpc is 
the depth over which light travels through the tricusp volume 
of the caustic ring in the tangent directions and $\sigma \sim 
D^2 \sim 2.5 \cdot 10^{-9}~{\rm cm}^2$ is the absorption 
cross-section.  This implies $n_d \sim 270/({\rm km})^3$.  
We take the cross-section for hard scattering of a dust 
particle on a dust particle to be of order 
$\sigma_d \sim 3~D^2 \sim 0.75 \cdot 10^{-8}~{\rm cm}^2$.  
Hence the mean free path between dust-dust scatterings 
$\lambda_d \sim {1 \over \sigma_d n_d} \sim$ 160 pc 
whereas the distance traveled transversely within 
the tricusp, where each flow is one of four different 
flows, is of order 100 pc, implying an optical depth 
for hard scattering of order 0.6.  That the optical 
depth is of order one provides a plausible explanation 
why the caustic imprints seen in the IRAS and Gaia maps 
are triangular rather than tricuspy.  Indeed, the 
trajectories forming the cusps encounter very high 
dust densities there. 

All trajectories through the tricusp pass through 
caustic folds, some trajectories participate in 
the formation of caustic folds, and among those 
some also participate in the formation of caustic 
cusps.  The fact of going through or participating 
in the formation of a fold does not increase the 
optical depth for scattering with other dust 
much because the density at a fold increases 
as $n \propto {1 \over \sqrt{h}}$ where $h$ 
is the distance to the fold surface and the 
${1 \over \sqrt{h}}$ singularity is integrable.  
On the other hand, the density diverges as 
${1 \over h}$ when a cusp is approached in 
the plane of the cusp \cite{sing}, leading to 
a logarithmic divergence in the optical depth 
for the flows forming the cusp.  Moreover, 
three flows participate in the formation of 
a cusp whereas only two flows participate in 
the formation of a fold.   Since the optical 
depth is order one, it is plausible that 
dust-dust collisions mess up the formation 
of cusps without messing up the formation 
of folds.  The sides of a triangular feature
associated with a caustic ring seen in a 
tangent direction indicate then the location 
of folds whereas the cusps are effectively 
erased.  A prediction of this interpretation 
is that the appearance of a tricusp imprint 
becomes more cuspy when the optical depth for 
obscuration by the dust in the feature is less, 
i.e. fainter caustic ring imprints will look 
more tricuspy.

\section{Big Flow properties}

We use the Gaia triangles to determine the properties of 
the four prominent flows on Earth associated with the nearby 
caustic.  Knowledge of their densities and velocities 
helps axion dark matter searches, particularly those using 
the cavity method \cite{cavity} and the echo method \cite{Arza}. 
We call the four flows Big, Little, Up and Down.  We estimate
the Sun's position relative to the nearby caustic ring by 
interpolating between the triangular features observed in 
the left and right tangent directions to the ring.  Then, 
from the Sun's position relative to the nearby caustic, we 
derive the flow velocities with respect to the Local Standard 
of Rest (LSR).  We also estimate the flow densities, and our 
errors on the various flow properties.    

From the triangles, we find that the caustic ring center is 
displaced from the Galactic Center by $5.8^\circ$ to the right. 
However, our distance from the caustic ring center $r_{\odot C}$ 
is not known.  If our distance to the inner tangent point is 
between 0.8 and 1.2 kpc (based on panel b of Fig.~9), then 
$r_{\odot C}$ has a value between 7.2 and 10.8 kpc. We assume 
$r_{\odot C}$ = 8.5 kpc in our estimates below. We also assume 
the velocity of the LSR to be $v_{\rm rot} = 220$ km/s in the 
direction of Galactic Rotation.  All our distances scale 
as $r_{\odot C}$, and all our velocities as $v_{\rm rot}$.  
For this reason, we do not quote any errors associated with 
imperfect knowledge of $r_{\odot C}$ and $v_{\rm rot}$. If 
$r_{\odot C}$ is found to differ from 8.5 kpc, one should 
simply multiply all our distances ($a$, $p$, $q$ ..) by 
$r_{\odot C}$/8.5 kpc.  Likewise for $v_{\rm rot}$.  Our 
estimates of the flow velocities are independent of $r_{\odot C}$.  
Our estimates of the flow directions are independent of both 
$r_{\odot C}$ and $v_{\rm rot}$ since they are determined by 
ratios of velocity components and each component scales as 
$v_{\rm rot}$.  There is however an error in the flow 
directions associated with uncertainty of the ratio 
$v/v_{\rm rot}$ where $v$ is the speed of the flows 
forming the 5th caustic ring.  According to the model 
\cite{Duffy}, $v$ = 520 km/s when $v_{\rm rot}$ = 220 km/s.  
$v$ = 520 km/s is the central value we use here.  From the 
success of the model in describing the pattern of caustic 
ring radii in the Milky Way \cite{MWcr}, we estimate that 
the error on $v/v_{\rm rot}$ is less than 3\%. 

\subsection{Previous estimates}

When only the left triangular feature in the IRAS sky map was 
known, the radius $a$ and width $p$ of the $5$th caustic ring 
were estimated \cite{MWcr} as: 
\begin{eqnarray}
a &\simeq& 8.31 \ \text{kpc} \nonumber \\
p &\simeq& 130 \ \text{pc} \ , 
\label{ap_old}
\end{eqnarray}
by assuming the ring to be axially symmetric and centered at the 
Galactic Center.  The Sun would then be outside the tricusp implying 
that there are two flows through our location, called Big and Little.  
Their densities and velocities were estimated to be~\cite{Duffy}: 
\begin{eqnarray}
d_+ &\simeq& 1.5 \times 10^{-24} \ \frac{\text{g}}{\text{cm}^3} \nonumber \\
d_- &\simeq& 0.15 \times  10^{-24} \ \frac{\text{g}}{\text{cm}^3} \nonumber \\
\vec{v} _\pm &\simeq& (\pm 120 \ \hat{\rho} + 505 \ \hat{\phi}) \ 
\frac{\text{km}}{\text{s}} \label{dv_old2}
\end{eqnarray}
where $\hat{\rho}$ points radially outward from the Galactic 
Center and $\hat{\phi}$ points in the direction of Galactic 
Rotation.  Because of an ambiguity in the sign of $\eta_0$, 
it is not clear which flow has the larger density.  If $\eta_0 < 0$, 
the flow of velocity $\vec{v}_\pm$ has density $d_\mp$.  If
$\eta_0 > 0$, the flow of velocity $\vec{v}_\pm$ has density $d_\pm$.

In Ref.~\cite{angmom} it was argued that formation of a Big Vortex 
in the axion dark matter fluid results in an enhanced dark matter 
density in the caustic rings. Any rotating BEC must have vortices.  
Whereas the vortices in superfluid $^4 \text{He}$ are repulsive, 
those in axion BEC are attractive.  Most of them merge to form a 
Big Vortex along the symmetry axis of the galaxy, enhancing the 
dark matter density in the Galactic Plane.  We assume this 
enhancement to be approximately by a factor 4 because this 
accounts for the prominence of the bumps in the Milky Way 
rotation curve attributed to caustic rings \cite{MWcr,angmom}.
The density estimates in Eq.~(\ref{dv_old2}) assume isotropic 
infall \cite{Duffy}.  Assuming the presence of a Big Vortex, 
they are modified to $d_+ \sim 6 \times 10^{-24}$ and 
$d_- \sim 0.6 \times 10^{-24}$ g/cc.  Let us emphasize
that densities near caustics have in any case large 
uncertainties because they are sensitive to position.

\subsection{The Sun's position relative to the nearby caustic}

The Galactic Coordinates $(l,b)$ of the vertices of the left 
triangle as observed in both the IRAS and Gaia sky maps are: 
$(77.86^\circ \pm 0.04 ^\circ , 3.3^\circ \pm 0.5 ^\circ)$, 
$(83.1^\circ \pm 0.4^\circ , 0.25^\circ \pm 0.15^\circ)$, 
$(77.85^\circ \pm 0.04 ^\circ , -2.4^\circ \pm 0.2 ^\circ)$. 
Those of the right triangle observed in the Gaia sky map are: 
$(-89.35^\circ \pm 0.05^\circ, 1.25^\circ \pm 0.05^\circ)$, 
$(-92.95^\circ \pm 0.05^\circ , -0.65^\circ \pm 0.05^\circ)$, 
$(-89.55^\circ \pm 0.05^\circ , -2.27^\circ \pm 0.03^\circ)$. 
The right triangle is located farther from the Galactic Center 
compared to the left triangle.  We interpret this to mean that 
the center of the $5$th caustic ring is displaced from the 
Galactic Center to the right as shown in Fig. 7. Assuming 
that the inner side of the caustic ring is an exact circle 
between the left and right tangent points, its center 
is located in the direction of Galactic Longitude 
$l = -5.80^\circ \ ^{+ 0.05^\circ} _{- 0.04^\circ}$. 
For the nominal value of our distance from the caustic 
ring center, $r_{\odot C}=8.5$ kpc, and assuming the 
caustic ring to be circular, its radius is 
$a = (8.448 \ ^{+ 0.001} _{-0.001})$ kpc. 
Assuming that the  caustic ring between the two tangent 
points lies in the plane determined by its center and the 
two midpoints of the inner edges of the triangles, we find 
that the caustic ring plane is tilted to the right by 
$\theta_t = 0.48^\circ \ ^{+ 0.20^\circ} _{-0.20^\circ}$ 
relative to the Galactic Plane.  We discuss below the 
errors due to failure of the assumption that the caustic
ring is exactly circular and planar between the two tangent
points, and include them in the error budget of the flow 
velocity vectors.

The location of the Sun relative to the caustic is specified 
by the horizontal and vertical sizes of the tricusp near the 
Sun ($p_\odot$ and $q_\odot$), the horizontal distance of the 
Sun ($x_\odot = r_{\odot C} - a$) from the inner side of the 
tricusp and its perpendicular distance ($z_\odot$) from the 
plane. We calculate the central values of these quantities 
based on the following assumptions: 1) the observed features 
are triangles inscribed by the tricusp; see Fig. 10,  2) the 
caustic ring is planar with its center displaced from the 
Galactic Center,  3) it has constant $a$, whereas $p$ and 
$q$ vary linearly between the left and right tangent 
directions.  The resulting central values and errors are: 
\begin{eqnarray}
p_\odot &=& (78.5 \ ^{+ 23.7} _{- 20.3}) \ {\rm pc} \ , \nonumber \\
q_\odot &=& (113.5 \ ^{+ 10.5} _{- 10.4}) \ {\rm pc} \ , \nonumber \\
x_\odot &=& (52.1 \ ^{+ 0.7} _{- 8.6}) \ {\rm pc} \ , \nonumber \\
z_\odot &=& (0.8 \ ^{+ 0.5} _{- 0.4}) \ {\rm pc} \ . \label{values_err}
\end{eqnarray}
We now briefly discuss the various sources of uncertainty. 

The largest source of uncertainty derives from ambiguity in estimating 
the horizontal size $p$ of the tricusp. If the observed triangles 
represent the cross-sections of the caustic ring near the inner 
tangent points, the linearly interpolated value of $p$ near the 
Sun is $p_\odot = ( 95.6 \ ^{+ 6.6} _{- 6.5} )$ pc where the 
uncertainties are associated with reading the vertices. On the 
other hand, if the outer vertices of the observed triangles tell 
us the directions of the outer tangent points, then $p_\odot = 
( 61.4 \ ^{+ 1.9} _{- 2.1} )$ pc.  We take the central value 
of $p_{\odot}$ to be $78.5$ pc.  The error on $p_\odot$ stated 
in Eq.~(\ref{values_err}) includes an additional contribution 
from the possible failure of the assumption of circularity of 
the ring, as discussed below.  The uncertainty on $q_\odot$ is 
much less than that on $p_\odot$ because it is determined by 
the inner side of the triangles. The uncertainty in $q_\odot$ 
is due to the errors associated with reading the vertices. 

For a planar caustic ring, the value of $z_\odot$ is $(0.8 \ ^{+ 0.5} 
_{-0.4})$ pc, including uncertainties from the vertices. We consider the 
case of a non-planar caustic ring where its outer cusp has height 
$z(\phi)  = A \cos \phi$ from a plane where $\phi$ is the 
azimuthal angle about the caustic ring center and $\phi = 0$ 
is the azimuth of the Sun.  The error in $z_\odot$ associated 
with non-planarity is less than $0.3$ pc if $A <$ 200 pc. 
$A >$ 200 pc seems unlikely since the caustic ring is seen  
close to the Galactic Plane in both tangent directions.  From 
a practical point of view, the errors in the flow velocities are 
dominated by the error in $p_\odot$ unless the error in $z_\odot$ 
is larger than $3$ pc. Such a large error in $z_\odot$ seems 
unlikely. 

For a constant ring radius $a$, we find $x_\odot = (52.1 \ ^{+0.7} 
_{-0.7})$ pc if only errors from reading the vertices are included.
Let us consider the possibility that the ring radius $a$ changes 
between the left and right tangent points as 
$a(\phi) = a_0 + a_1 \phi + a_2 \phi ^2$ where $\phi$ varies from 
$-6.3^\circ$ to $+6.3^\circ$ between the two inner tangent directions, 
and $a_0$ = 8.448 kpc.  When $|a_1|$ is increased, one tangent point 
comes closer to the Sun whereas the other moves away.  We require 
$|a_1|$ to be less than 314.5 pc so that the distance to the nearest 
tangent point remains larger than half the distance in the constant 
radius case.  Furthermore we assume that the second order coefficient 
$|a_2| \lesssim |a_1|/2$. Then, $x_\odot$ is found to range between 
$43.5$ and $51.6$ pc, and the value of $p_\odot$ between $58.2$ and 
$94.9$ pc including the uncertainty discussed in the paragraph before 
last.  

We find that the Sun is almost certainly within the tricusp volume
of the caustic ring.  Given $x_\odot < p_\odot$, whether the Sun is 
inside or outside the tricusp is determined by its vertical distance  
$z_{\odot}$ from the caustic ring plane. For the central values of 
$x_\odot , p_\odot$ and $q_\odot$, the Sun is outside the tricusp 
if $z_\odot \geq 7.0$ pc which is very unlikely according to our 
estimates.  However, the Sun is outside the tricusp for some 
extreme values of the parameters, e.g. $p_\odot = 58.2$ pc, 
$x_\odot = 52.8$ pc, $z_\odot = 1.3$ pc and all plausible 
values $q_\odot$.  We estimate the probability that the Sun 
is outside the tricusp to be less than 1\%.   Assuming the 
Sun is indeed inside the tricusp, there are four prominent 
flows on Earth associated with the nearby caustic ring.  In 
Fig. 11, for various values of $z_\odot$, we show the directions 
of the flows with respect to the LSR in the $\eta_0 < 0$ case
and indicate their densities by the sizes of circles. As the 
Sun moves closer to the boundary of the tricusp, two flows 
approach each other in velocity space while their densities 
increase.  They disappear the moment the Sun passes outside 
the tricusp.

\subsection{Big Flow velocity vector and density estimates}

The flow velocities in the frame of the caustic ring are calculated 
using  Eqs.~(\ref{GAIA_XT}) and (\ref{GAIA_vrhozphi}).  The caustic 
ring parameters near the location of the Sun are derived from the 
central values of $a$, $p_\odot$ and $q_\odot$ and setting 
$v = 520$ km/s as the speed of the flow \cite{Duffy}:
\begin{eqnarray}
u &=& \frac{v^2 - v_{\text{rot}}^2 }{a} = 26.3 \times 10^3 \ 
{\rm kpc}^{-1} {\rm (km/s)}^2 \nonumber \\
\eta_0 &=& \pm \sqrt{\frac{2p}{u}} = \pm 2.44 \times 10^{-3} \ 
{\rm kpc} {\rm (km/s)}^{-1} \nonumber \\
\zeta &=& \frac{27}{16} \frac{p^2}{q^2} = 0.807~~\ .
\label{parameters_est}
\end{eqnarray}
Eqs.~(\ref{GAIA_vrhozphi}) give the components of the 
flow velocities in cylindrical coordinates attached to the 
caustic ring. Since the caustic ring center is displaced 
from the Galactic Center by $\theta = (5.80^\circ \ ^{+
0.05^\circ} _{- 0.04^\circ} )$ to the right, and the 
caustic ring plane is tilted relative to the Galactic
Plane by $\theta_t = (0.48^\circ \ ^{+ 0.20^\circ}
_{-0.20^\circ} )$ also to the right, the velocity components 
$(v_{j \rho}^{\rm G}, v_{j \phi}^{\rm G}, v_{jz}^{\rm G})$
of the jth flow in Galaxy centered cylindrical 
coordinates are obtained from the components 
$(v_{j \rho}, v_{j \phi}, v_{jz})$ in caustic
centered cylindrical coordinates using
\begin{eqnarray}
v_{j \rho} ^{\rm G} &=& \cos \theta \ v_{j\rho} - \sin \theta \ \cos 
\theta_{\rm t} \ v_{j\phi} + \sin \theta \ \sin \theta_{\rm t} \ v_{j z} 
\nonumber \\
v_{j \phi} ^{\rm G} &=& \sin \theta \ v_{j\rho} + \cos \theta \ \cos 
\theta_{\rm t} \ v_{j\phi} - \cos \theta \ \sin \theta_{\rm t} \ v_{j z} 
\nonumber \\
v_{j z} ^{\rm G} &=& \sin \theta_{\rm t} \ v_{j\phi} + 
\cos \theta_{\rm t} \ v_{j z}  
\label{GAIA_vcomps}
\end{eqnarray}
where $\hat{\rho} _{\rm G}$ points away from the Galactic Center, 
$\hat{\phi} _{\rm G}$ points in the direction of Galactic Rotation 
and $\hat{z} _{\rm G}$ points to the Galactic North Pole. The flow 
velocity with respect to the LSR is  
$\vec{v}_{j {\rm LSR}} = \vec{v}_{j} ^{\rm G} 
- v_{\rm rot} \ \hat{\phi} _{\rm G}$ 
with $v_{\rm rot} = 220$ km/s. The error associated with uncertainty
on $v/v_{\rm rot}$ enters here.  It contributes of order $0.25^\circ$
to the errors on the velocity directions, given in Eqs.~(\ref{veldir})
and (\ref{veldir2}) below. 

Table \ref{GAIA_table_1} gives the densities $d_j$ and velocities 
$\vec{v} _j ^{\rm G}$ of the four flows corresponding to the 
central values of $x_\odot$, $z_\odot$, $p_\odot$ and $q_\odot$ 
when $\eta_0 < 0$. Table \ref{GAIA_table_2} gives the same 
information in case $\eta_0 > 0$.  The densities are calculated 
using Eqs.~(\ref{density_formula}).  We set $b = v = 520$ km/s.  
We do not have enough information to determine $b$ precisely 
but it is expected to be of order $v$ \cite{sing}.  The 
difference between $b$ and $v$ is relatively unimportant 
in view of the other uncertainties affecting the densities.  
For the infall rate, in view of the Big Vortex, we multiplied 
by 4 the estimate given in ref. \cite{Duffy}, 
i.e. ${dM \over d\Omega d\eta} = 4 \times 
{7.8~M_\odot \over {\rm sterad}~{\rm yr}}$.  Estimates of the 
densities are very uncertain because the densities vary rapidly 
with  position.  Over the range of plausible parameter values, 
Eqs.~(\ref{values_err}), the Big and Up flows range from 1/2 
their central values to infinity (when the Sun approaches the 
tricusp boundary),  the Down flow ranges from 1/2 to 4 times 
its central value, and the Little flow changes by 20\%.  All 
entries in Tables I and II are highly correlated since they 
are functions of a small number of parameters, mainly $x_\odot$, 
$p_\odot$, $q_\odot$ and $z_\odot$.

The directions of the flow velocities with respect to the LSR are:
\begin{eqnarray}
(l,b)|_{\rm Big} &=& (70.17^\circ \ ^{+0.84^\circ} _{-0.19^\circ} , 
1.14^\circ \ ^{+2.09^\circ} _{-0.59^\circ}) \nonumber\\
(l,b)|_{\rm Little} &=& (89.96^\circ \ ^{+0.07^\circ}_{-0.87^\circ} 
, 0.86^\circ \ ^{+0.38^\circ} _{-0.36^\circ}) \nonumber\\
(l,b)|_{\rm Up} &=& (67.97^\circ \ ^{+1.91^\circ}   
_{-1.81^\circ} , 8.28^\circ \ ^{+2.44^\circ} _{-5.01^\circ}) \nonumber\\
(l,b)|_{\rm Down} &=& (67.81^\circ \ ^{+1.72^\circ} _{-1.76^\circ} , 
-7.05^\circ \ ^{+3.58^\circ} _{-2.35^\circ})~~\ .
\label{veldir}
\end{eqnarray}
in case $\eta_0 < 0$ and 
\begin{eqnarray}
(l,b)|_{\rm Big} &=& (89.96^\circ \ ^{+0.19^\circ} _{-0.87^\circ} ,
0.49^\circ \ ^{+0.61^\circ} _{-2.15^\circ}) \nonumber\\
(l,b)|_{\rm Little} &=& (70.17^\circ \ ^{+0.84^\circ}_{-0.06^\circ}
, 0.77^\circ \ ^{+0.36^\circ} _{-0.36^\circ}) \nonumber\\
(l,b)|_{\rm Up} &=& (92.40^\circ \ ^{+1.82^\circ}
_{-1.78^\circ} , 8.91^\circ \ ^{+2.43^\circ} _{-3.69^\circ}) \nonumber\\
(l,b)|_{\rm Down} &=& (92.18^\circ \ ^{+1.85^\circ} _{-1.93^\circ} ,
-6.90^\circ \ ^{+5.20^\circ} _{-2.52^\circ}) .
\label{veldir2}
\end{eqnarray}
in case $\eta_0 > 0$.  The directions and errors are displayed 
in Figs. 12 and 13 for $\eta_0 < 0$ and $\eta_0 > 0$ respectively.  
To obtain the flow velocities with respect to an observer on 
Earth, one needs to subtract from the velocities given in 
Tables I and II the velocity of the Sun with respect 
to the LSR and the velocity of the observer with respect 
to the Sun due to the orbital and rotational motions of 
the Earth.

\begin{table*}
\caption{\label{GAIA_table_1}Central values of the densities 
and velocities of the flows through the Sun associated with the 
nearby caustic in the Galactic Rest Frame in case $\eta_0<0$.  
The densities are uncertain by a factor of 2 or more, as 
discussed in the text.  The error on the velocity components 
is dominated by the uncertainty in the rotation speed of the 
LSR, taken to be 220 km/s but known only within approximately 
10\%.  The velocity directions and their errors are given 
explicitly in Eqs.~(\ref{veldir}) and Fig. 12.}

\begin{ruledtabular}
\begin{tabular}{ccccc}
Flow & $d$ [$10^{-24}$ g/cm$^3$] & $v_{\rho} ^{\rm G}$ [km/s] & $v_{\phi} 
^{\rm G}$ [km/s] & $v_{z} ^{\rm G}$ [km/s] \\
\hline
Big & $20.0$ & $-104.4$ & $509.4$ & $6.1$ \\
Little & $2.0$ & $-0.2$ & $520.0$ & $4.5$ \\  
Up & $9.6$ & $-115.3$ & $505.1$ & $44.8$ \\
Down & $8.4$ & $-116.4$ & $505.4$ & $-38.1$ \\ 
\end{tabular}
\end{ruledtabular}
\end{table*}

\begin{table*}
\caption{\label{GAIA_table_2} Same as Table \ref{GAIA_table_1}
but for $\eta_0 > 0$. The velocity directions and their
errors are given explicitly in Eqs.~(\ref{veldir2}) and Fig. 13.}

\begin{ruledtabular}
\begin{tabular}{ccccc}
Flow & $d$ [$10^{-24}$ g/cm$^3$] & $v_{\rho} ^{\rm G}$ [km/s] & $v_{\phi} 
^{\rm G}$ [km/s] & $v_{z} ^{\rm G}$ [km/s] \\
\hline
Big & $20.0$ & $-0.2$ & $520.0$ & $2.7$ \\    
Little & $2.0$ & $-104.4$ & $509.4$ & $4.4$ \\     
Up & $8.4$ & $12.7$ & $517.7$ & $47.1$ \\
Down & $9.6$ & $11.5$ & $518.6$ & $-35.8$ \\ 
\end{tabular}
\end{ruledtabular}
\end{table*}

\section{Summary}

In this paper we added to the observational evidence in 
support of caustic rings and the Caustic Ring Model.  The 
additional evidence is found in the rotation curve of our 
closest large Galactic Neighbor M31 \cite{Chemin} and in a 
triangular feature in the Gaia map \cite{Gaia1,Gaia2} of the 
Milky Way in the direction of Galactic Coordinates $(l,b) = 
(-91^\circ, 0^\circ)$.  The M31 rotation curve has bumps 
whose locations are in rough agreement with the model 
predictions for the radii of the first three ($n=1,2,3$) 
caustic rings.  The bumps attributed to the $n=1$ ring are 
particularly striking as they appear at the same location 
on both the receding and approaching sides of the rotation 
curve, strongly suggesting the existence of a ring structure 
in the M31 halo with a diameter of order 60 kpc.

The Gaia triangular feature at $(l,b) = (-91^\circ, 0^\circ)$
solves a question raised by a previous claim \cite{MWcr}
that a triangular feature in the IRAS map, in the direction 
$(l,b) = (80^\circ,0^\circ)$, is the imprint of the 5th 
caustic ring ($n=5$) on baryonic matter in the Galactic 
Disk, seen in a tangent direction to the caustic ring
from our viewpoint.  The 5th caustic ring has two tangent 
directions from our viewpoint, and there is no triangular 
feature in the IRAS map on the opposite side, near 
$(l,b) = ( - 80^\circ, 0^\circ)$.  In contrast, the 
Gaia map has two triangular features at nearly symmetrical 
locations, one which coincides with the IRAS feature on 
the left side, and the new feature on the right side.  
The triangular feature on the right side, like the one 
on the left, is to a high level of accuracy an isosceles 
triangle with axis parallel to the Galactic Plane.  It 
has approximately the same aspect ratio as the left 
triangle but is 40\% smaller.  It was emphasized earlier 
\cite{Natarajan} that the transverse size of a caustic 
ring cross-section may vary along the ring.

The Gaia triangles are features in the distribution 
of dust. Like the IRAS triangle, they are much sharper 
than they would be if produced by gas and dust in thermal 
equilibrium in the gravitational field of the caustic 
ring \cite{starsgas}.  We propose here instead that dust 
is entrained by the axion flows forming the 5th caustic 
ring.  By following the same trajectories as the axions, 
the dust particles form the same caustics. This would 
explain the sharpness of the features.  Section III 
discusses the entrainment of a dust particle by a 
cold axion flow.  The axions in the flow are assumed 
to be a highly degenerate Bose gas at temperature $T$, 
with almost all axions therefore in a single state 
forming a Bose-Einstein condensate.   We derived a 
formula, Eq.~(\ref{dec}), for the frictional deceleration 
of a dust particle moving with respect to the cold axion 
flow.  The deceleration is proportional to $T^{-1}$.  We 
estimate the temperature cold dark matter axions have 
today taking account of the heat in the axions themselves 
and of heat they absorb by cooling baryons.  Our formula 
for friction indicates that the dust particle is efficiently 
entrained by the axion flow. However, in the Galactic Disk 
collisions with gas may slow down the dust particle 
considerably.  On the other hand the collisions with 
gas do not diffuse the flow of dust particles.  We 
conjecture that the dust particles, although slowed 
down by collisions with gas in the disk, follow the 
same trajectories as the axions and form the same 
caustics.   

We also proposed an explanation why the features seen 
in the IRAS and Gaia maps are triangular even though 
the cross-section of a caustic ring is tricuspy.  For 
accepted values of the dust density and dust grain 
size, the flow of dust inside the caustic ring is 
collisionless but only barely so.   The density in 
the three flows forming the cusps of a tricusp is much 
higher than elsewhere within the tricusp. Dust-dust 
collisions within the cusps are likely to fuzz them 
up.  The observed features are then qualitatively 
triangles inscribed in the tricusp, as indicated 
in Fig. 10.  Finally, we proposed an explanation 
why no right triangle is seen in the IRAS map.  
IRAS observes in the infra-red.  Dust particles 
emit in the infrared when heated by stellar radiation.  
In the left tangent direction, the 5th caustic ring 
lies in the midst of a spiral arm with abundant 
stellar activity whereas in the right tangent 
direction, the ring lies in a quiet region. The 
dust on the right sides receives less heat and 
for this reason fails to show up in the IRAS map.

The GAIA and IRAS triangular features imply that the 
caustic ring center is displaced from the Galactic
Center to the right by $5.8^\circ$, and that the 
plane of the ring is tilted relative to the Galactic 
Plane $0.48^\circ$ to the right.  In all likelihood, 
we on Earth are inside the tricusp volume of the 
nearby caustic ring.  As a result there are four 
flows on Earth associated with the nearby caustic, 
called Big, Little, Up and Down.   In contrast, on 
the basis of pre-Gaia observations when only the IRAS 
triangle was known and the caustic ring center was 
assumed to coincide with the Galactic Center, it was 
thought that we are outside the tricusp and hence 
that there are two flows on Earth associated with 
the nearby caustic, Big and Little.  Being inside
the tricusp implies two additional flows, Up and 
Down.

The flows that form the 5th caustic ring are prominent
on Earth.  They produce narrow peaks in the axion energy 
spectrum which are observable with great resolution in 
the cavity haloscopes \cite{cavity}.  These detectors 
are made more sensitive by searching for such narrow 
peaks \cite{Hires}. Knowing the velocity vectors of 
the flows on Earth one can calculate how the peaks 
move as a function of time of day and time of year 
\cite{Ling}, so that observations made at different 
times can be related to one another.  Generally speaking, 
axion dark matter searches are helped by knowing the 
velocity distribution of the axions on Earth.  In 
particular, a recently proposed axion dark matter 
detection scheme \cite{Arza}, called the axion echo 
method, is largely predicated on the existence of one 
or more cold flows and on knowledge of their velocity 
vectors.  So there is strong motivation to determine the 
velocity vectors on Earth of the flows forming the nearby 
caustic ring.  This is our purpose in Section IV.  Even 
accepting that the IRAS and Gaia triangles show the imprints 
of the 5th caustic ring in the two tangent directions, 
uncertainties arise because of the need to interpolate 
between the two tangent points, which are 940 pc away 
from us on either side.  To find the central values of 
the flow velocity vectors we assume that the ring between 
the two tangent points is planar and circular.  We also 
assume that the caustic ring transverse sizes at our 
location, $p$ and $q$, can be obtained from estimates at 
the tangent points by linear interpolation.  We estimate 
the likely errors associated with these assumptions.  The 
results are given in Eqs.~(\ref{veldir}) and (\ref{veldir2}), 
Tables I and II, and Figs. 12 and 13.  These predictions 
are testable as soon as axion dark matter is detected in 
the laboratory, and perhaps by other means.

\begin{acknowledgments}

We gratefully acknowledge stimulating discussions with 
Peter Barnes, David Tanner, Shriram Sadashivajois, Ariel Arza
and Richard Bradley.  This work was supported in part by the 
U.S. Department of Energy under grant DE-SC0010296 at the 
University of Florida. SSC is supported by the grant 
``The Milky Way and Dwarf Weights with Space Scales'' 
funded by University of Torino and Compagnia di 
S. Paolo (UniTO-CSP). SSC also acknowledges partial 
support from the INFN grant InDark and the Italian 
Ministry of Education, University and Research (MIUR) 
under the Departments of Excellence grant L.232/2016.

\end{acknowledgments}

\appendix

\section{Statistical significance of the Gaia left 
and right triangles}

The Gaia skymap is a greyscale map of the logarithm $g(l,b)$ 
of the number of stars per unit solid angle \cite{Gaia2}
observed by the Gaia satellite in the direction of Galactic
Coordinates $(l,b)$.  We obtained $g(l,b)$ from the Gaia
data archive \cite{Gaiaweb} over a grid of spacing
$\delta l = \delta b = 0.1^\circ$ covering the range      
$- 180^\circ < l < 180^\circ$ and $-5^\circ < b < 5^\circ$.
We made a cut on the distances from Earth to the stars in the 
Gaia catalog since the signal is due only to stars that are 
behind the caustic ring tangent points, approximately 1 kpc 
away. We only used sources that are more than 0.8 kpc away from 
Earth.

First, we convoluted the map with a square top hat filter
of size $\Delta l = \Delta b = 5^\circ$ placed at 72 = 360/5
successive adjacent positions along the Galactic Plane.  
For each position, we calculated the average value $\bar{g}_i$
($i = 1, 2, ..., 72)$ of $g(l,b)$ within the rectangle at that
position, and the deviation $\Delta \bar{g}_i$ of $\bar{g}_i$
from its average over the $n$ neighboring positions to the
left plus the $n$ neighboring positions to the right, where
$n$ = 5, 3 and 2.  Changing the value of $n$ did not change
the final results significantly.  The results reported here 
are for  $n = 5$.  The three largest negative fluctuations 
in $\Delta \bar{g}_i$ are at the positions of the right 
($l \simeq - 91^\circ$) and left ($l \simeq 80^\circ$) 
triangles and at a relatively dark area in the Gaia skymap 
near ($b \simeq 0, l \simeq 37^\circ$).  These three
largest negative fluctuations are respectively
- 2.5 $\sigma$, - 3.8 $\sigma$ and - 2.6 $\sigma$
where $\sigma$ is the root mean square of all the 
$\Delta \bar{g}_i$.  Indeed, looking by eye, the 
darkest areas along the Galactic Plane are seen at 
those three locations in the Gaia skymap.  The dark area 
at $l \simeq 37^\circ$ is somewhat triangular in shape,
pointing to the left like the left triangle. Its position
is consistent with that of a prominent rise in the inner 
Galactic Rotation Curve derived from the Massachusetts-Stony 
Brook North Galactic Plane CO Survey \cite{CO} between 
$35.3^\circ$ and $38.7^\circ$ and attributed to the 9th 
caustic ring in the Caustic Ring Model of the Milky Way 
halo \cite{MWcr}.  As the 9th caustic ring is not the 
topic of this paper, we do not discuss the dark area
near $l \simeq 37^\circ$ further.  We remove it from 
the data so that it is counted neither as signal nor 
as noise.

Second, we convoluted the map with a triangular top hat 
filter whose size and Galactic Latitude matches that of 
the right triangle in the Gaia skymap.  It is isosceles, 
parallel to the Galactic Plane, pointing to the right, with 
height $\Delta b = 3.6^\circ$, width $\Delta l = 3.5^\circ$ 
and displaced $0.5^\circ$ below the Galactic Plane.  These 
properties are consistent with the positions of the 
vertices of the right triangle stated in Section 4.B. 
The filter is moved to the left and right and the 
average $\bar{g}(l)$ of $g(l,b)$ over the area of 
the filter is calculated for each position $l$. We 
then fit $\bar{g}(l)$ with a second order polynomial in 
$l$ after removing the regions $(73^\circ, 85^\circ)$, 
$(33^\circ, 45^\circ)$, and $(-95^\circ, -85^\circ)$ 
which contain the features attributed to caustic rings.
We also exclude $(-75^\circ, -45^\circ)$ as it is along the 
direction of a major spiral arm of the Milky Way. The deviation 
$\Delta \bar{g} \equiv \bar{g} - \bar{g}_{\rm fitted}$ from the 
fitted curve is $-5.1 \,\sigma$ where $\sigma$ is the root mean 
square of all $\Delta\bar{g}$ other than those in the excluded 
regions.  This is illustrated in the top panel of Fig. 14.

Finally, we convoluted the map with a triangular top hat
filter whose size and Galactic Latitude matches that of 
the left triangle in the Gaia skymap.  It is isosceles, 
parallel to the Galactic Plane, pointing to the left, with 
height $\Delta b = 5.7^\circ$, width $\Delta l = 5.2^\circ$ 
and displaced $0.4^\circ$ above the Galactic Plane.  These
properties are consistent with the positions of the
vertices of the left triangle stated in Section 4.B.
The analysis is similar to that using the right triangle 
top hat filter. The filter is moved to the left and 
right and the average $\bar{g}(l)$ of $g(l,b)$ over 
the area of the filter is calculated for each position 
$l$.  $\bar{g}(l)$ is fitted with a second order 
polynomial in $l$ after exclusion of the same regions 
as stated above.  The deviation 
$\Delta \bar{g} \equiv \bar{g} - \bar{g}_{\rm fitted}$ 
at the position of the left triangle is 
$-6.5 \,\sigma^\prime$ where $\sigma^\prime$ is the 
root mean square of all $\Delta\bar{g}$ other than 
those in the excluded regions.  This is illustrated
in the bottom panel of Fig. 14.

\clearpage

\begin{figure}
\begin{center}
\includegraphics[height=110mm]{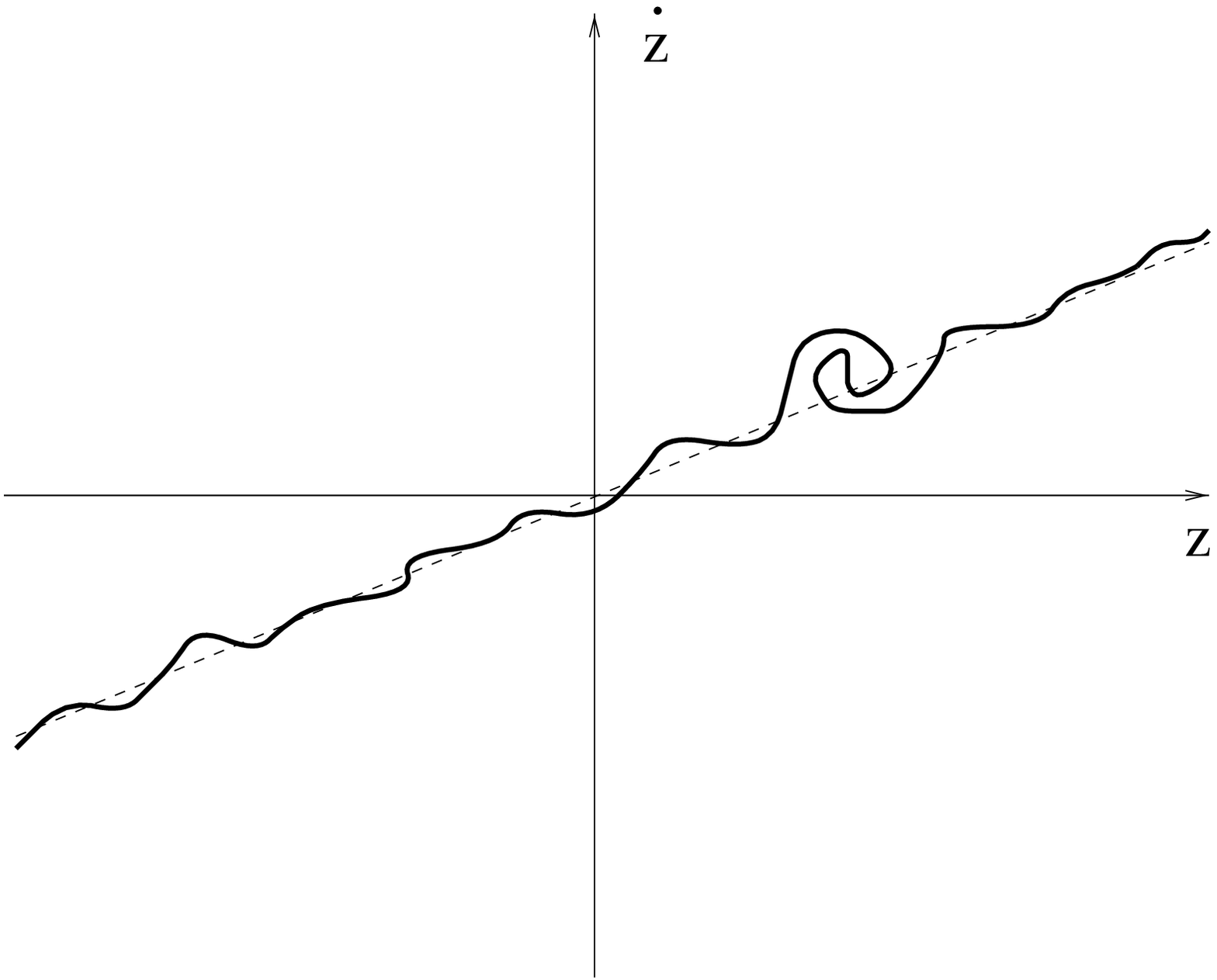}
\vspace{0.3in}
\caption{Illustration of the behaviour of the 3-dimensional 
hypersurface, called the phase-space sheet, on which cold 
collisionless dark matter particles lie in 6-dimensional 
phase-space.  The wiggly line is the intersection of the 
phase-space sheet with the $(z,\dot{z})$ plane.  The 
thickness of the line is the primordial velocity 
dispersion.  The broad slope of the line is the 
Hubble-Lemaitre expansion rate.  The wiggle amplitudes
are the local peculiar velocities.  Where an overdensity 
grows in the non-linear regime, the line winds up in 
clockwise fashion.  One such overdensity is shown.
The figure is taken from ref.~\cite{sing}.
Reproduced with permission of the journal.}
\end{center}   
\label{fig1}
\end{figure}

\begin{figure}
\begin{center}
\includegraphics[height=110mm]{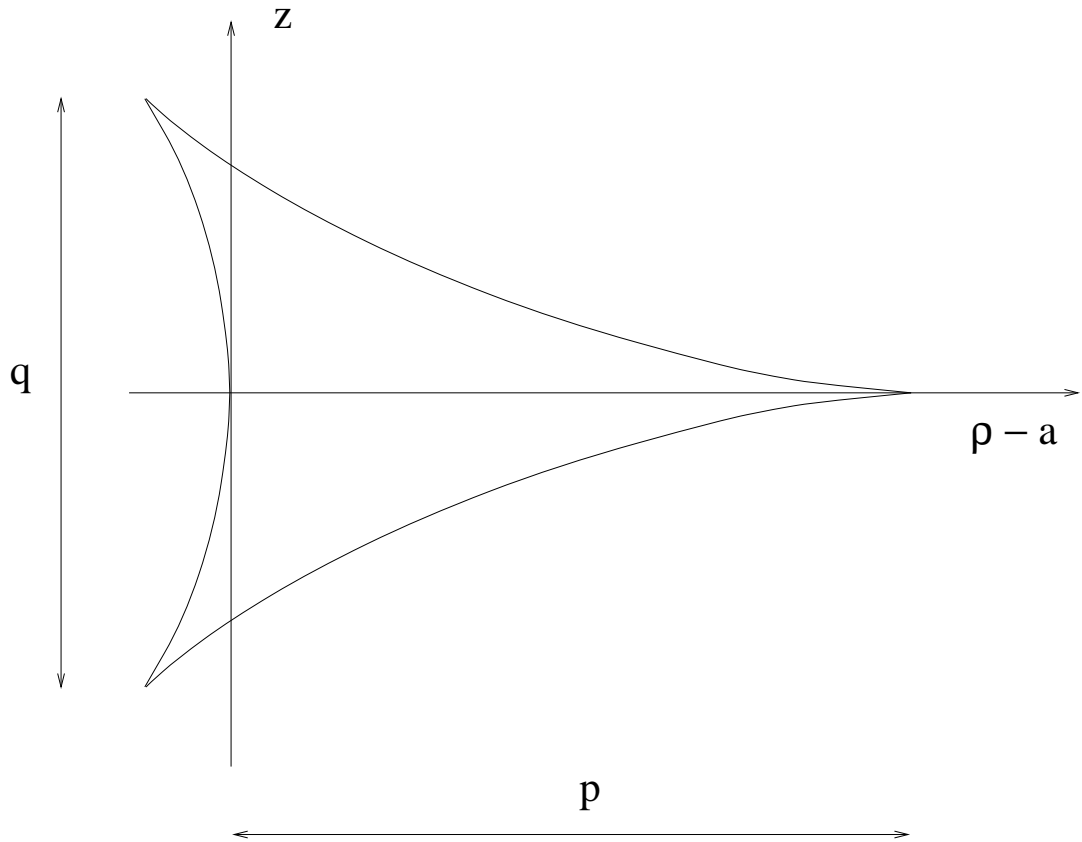}
\vspace{0.3in}
\caption{The tricusp cross-section of a caustic ring on
the $\rho$-$z$ plane. The dimensions of the tricusp in
the radial and vertical directions are called $p$ and $q$,  
respectively.  The figure is taken from ref.~\cite{sing}.
Reproduced with permission of the journal.}
\end{center}
\label{tricusp}
\end{figure}

\begin{figure}
\begin{center}
\includegraphics[height=130mm]{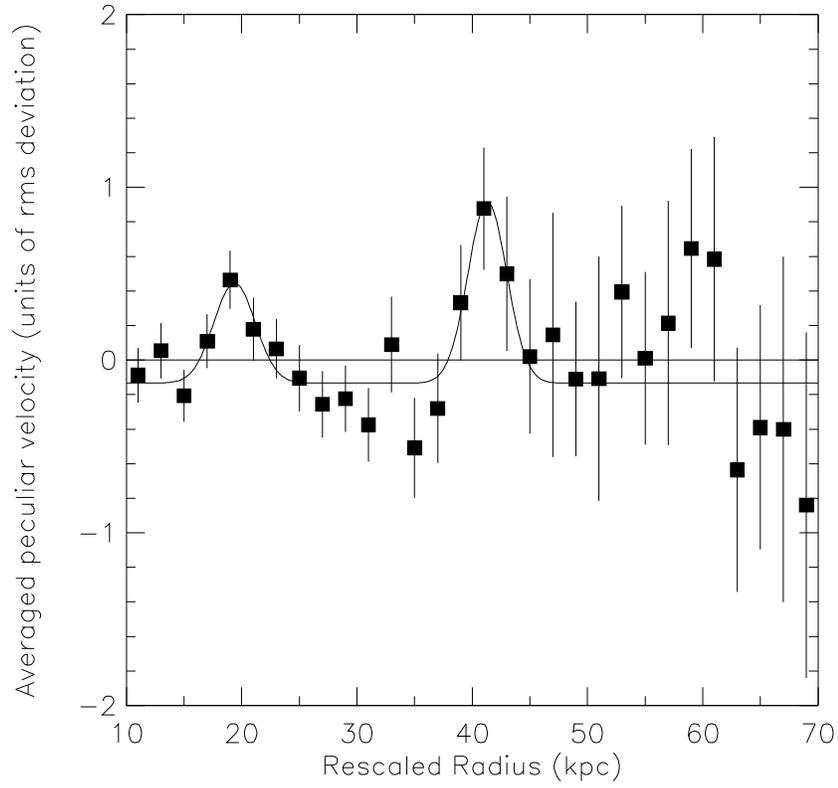}
\vspace{0.3in} 
\caption{Combined rotation curve of a sample of 
32 galaxies, as described in the text.  The figure 
is taken from ref. \cite{Kinney}.  Reproduced with 
permission of the journal.}
\end{center}
\label{32combined}
\end{figure}

\begin{figure}
\begin{center}
\includegraphics[height=100mm]{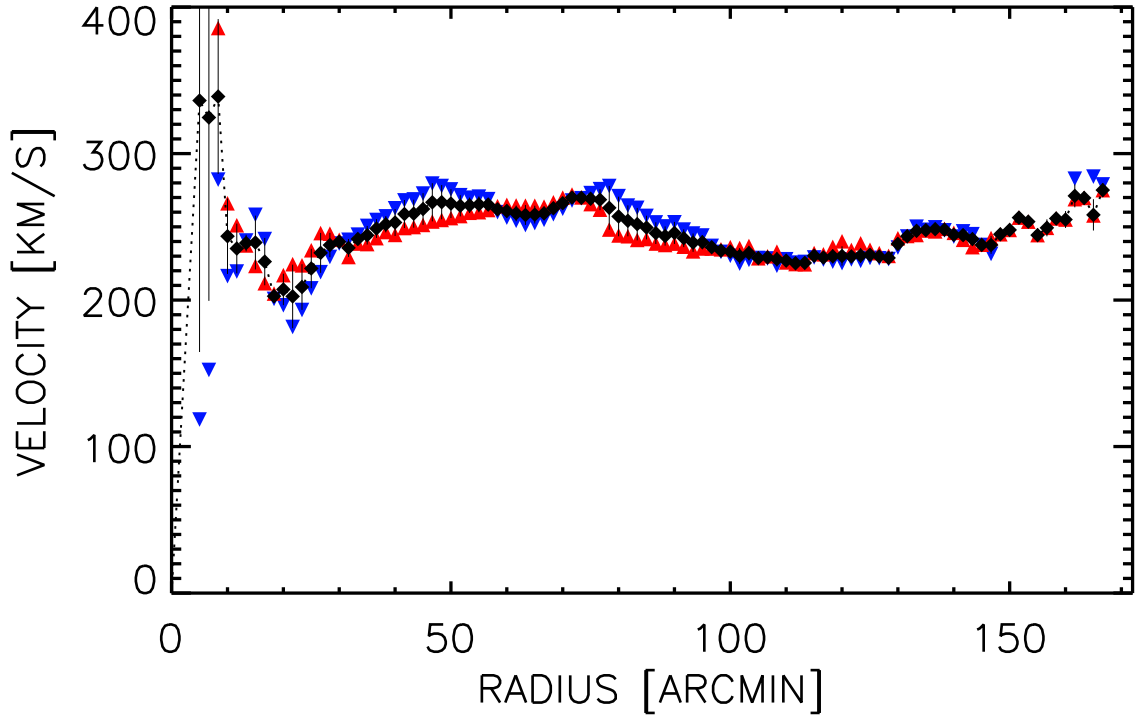}
\vspace{0.3in}
\caption{Rotation curve of M31 obtained by L. Chemin, 
C. Carignan and T. Foster \cite{Chemin}. Filled  diamonds
are  for  both halves  of  the  disc  fitted  simultaneously  
while  blue downward/red upward triangles are for the 
approaching/receding sides fitted separately.  The bumps 
mentioned in the text are near 135 arcmin = 30.9 kpc,
70 arcmin = 16 kpc  , and 47 arcmin = 10.8 kpc.  
Reproduced with permission from the journal.}
\end{center}
\label{M31}
\end{figure}

\begin{figure}
\begin{center}
\includegraphics[height=100mm]{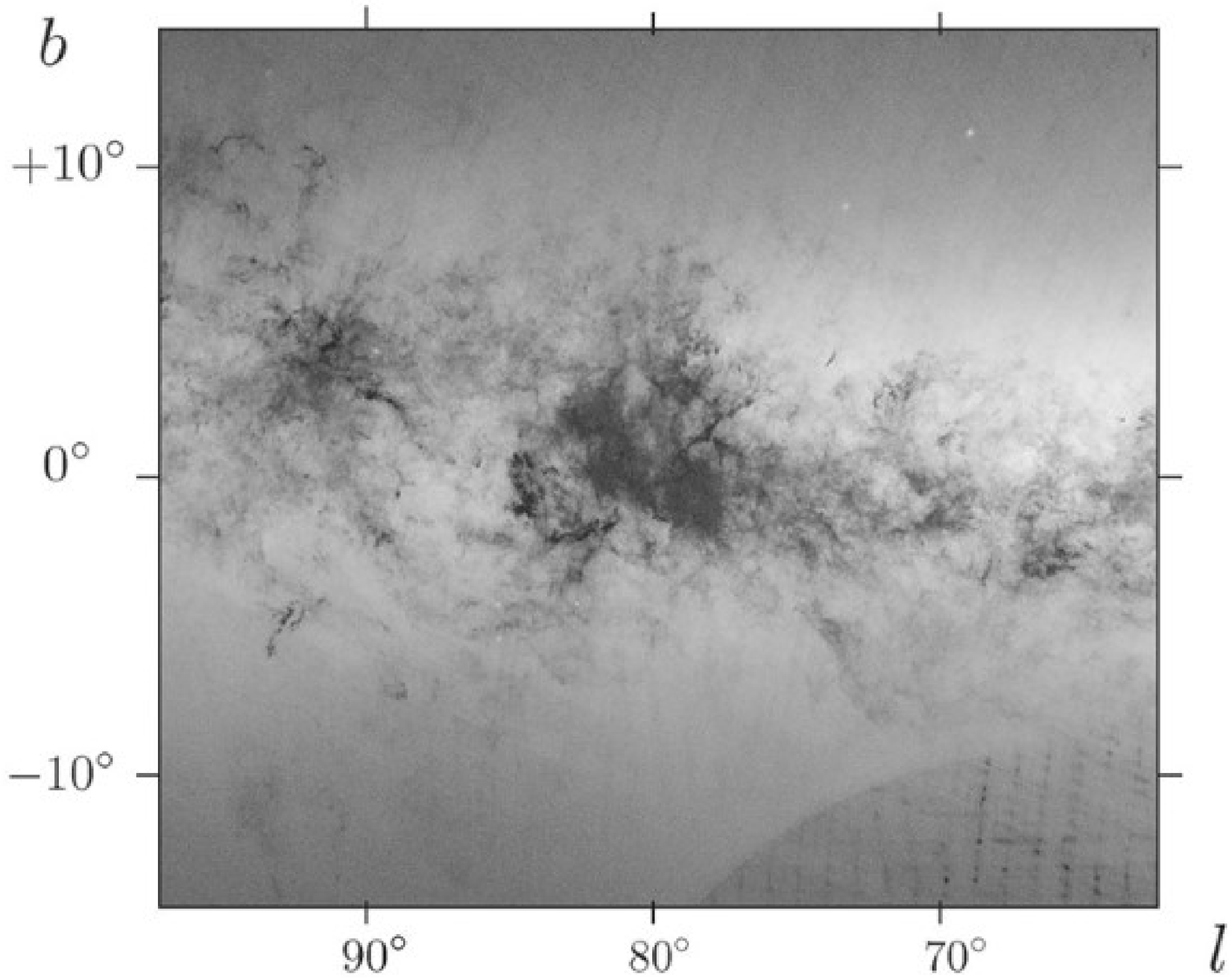}
\vspace{0.3in}
\caption{Part of the Gaia skymap of the Galactic 
Plane with the left triangular feature at the center.}
\end{center}
\label{lgaia}
\end{figure}

\begin{figure}
\begin{center}
\includegraphics[height=100mm]{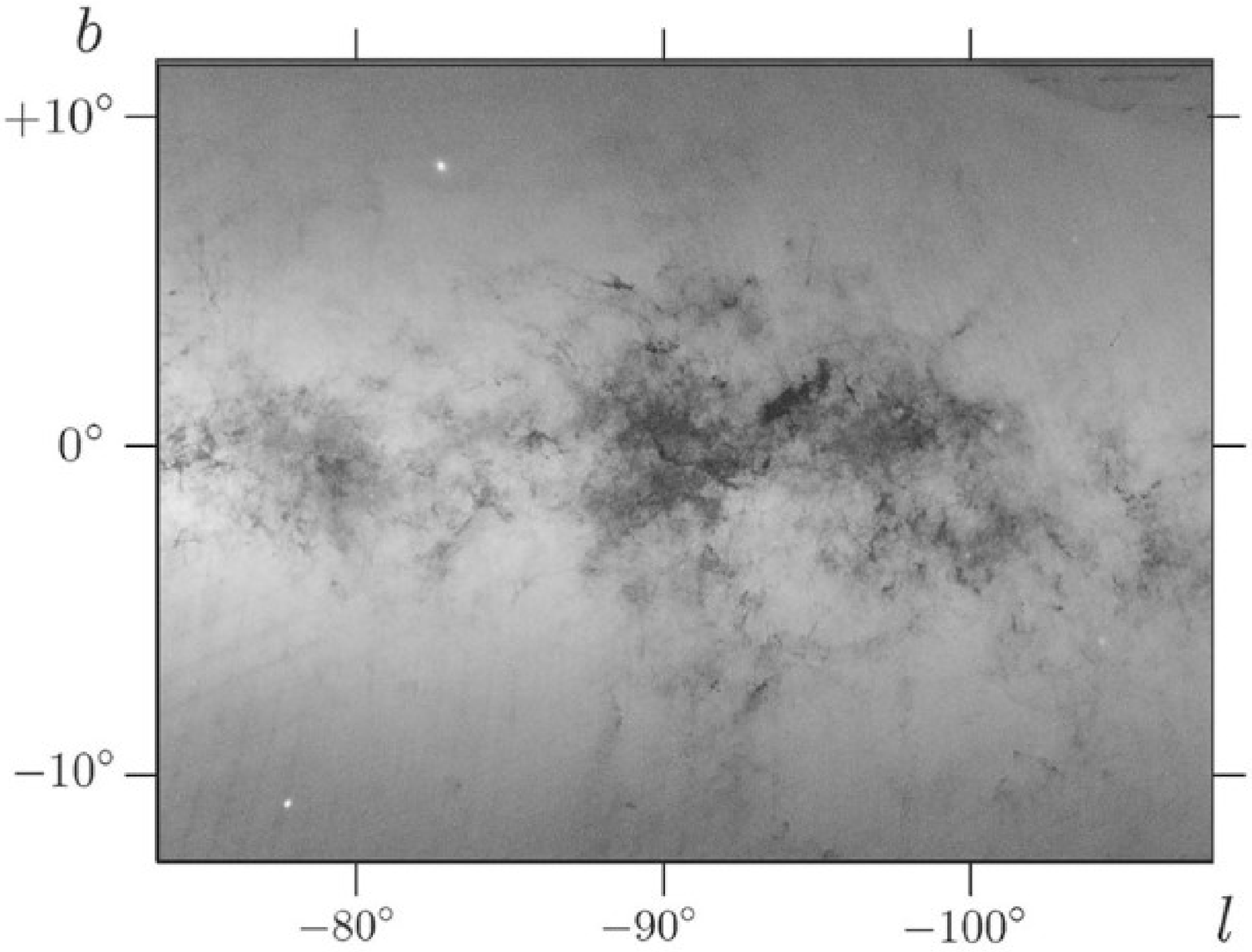}
\vspace{0.3in}
\caption{Part of the Gaia skymap of the Galactic
Plane with the right triangular feature at the center.}
\end{center}  
\label{rgaia}
\end{figure}

\begin{figure}
\begin{center}
\includegraphics[height=100mm]{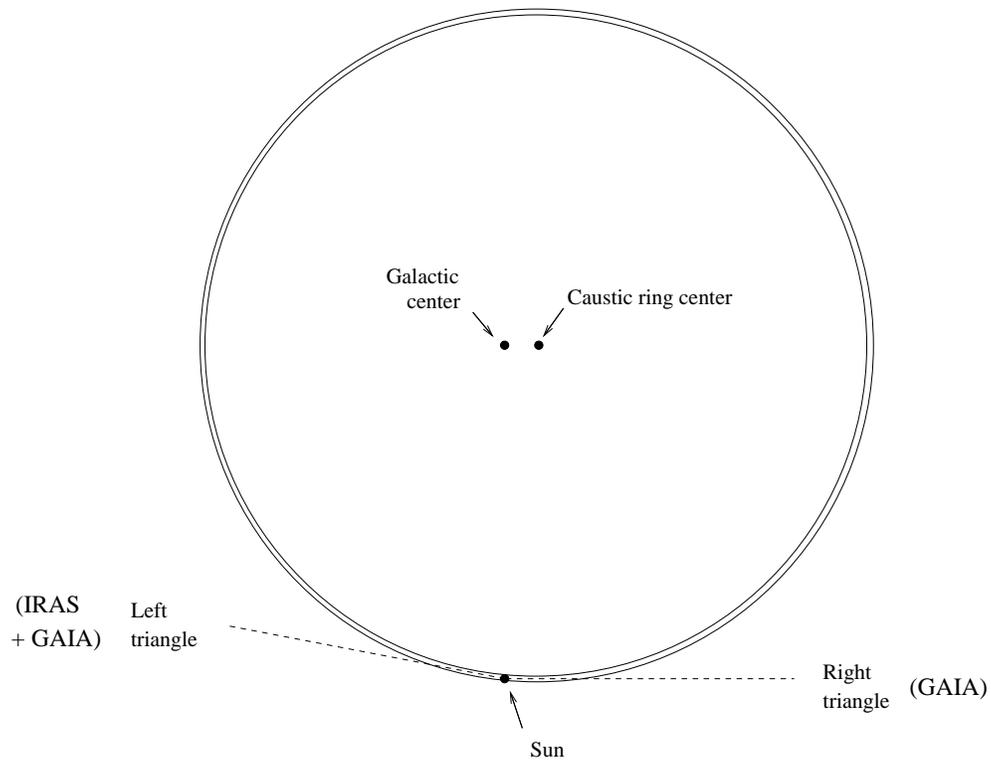}
\vspace{0.3in}
\caption{Illustration of the relative positions of 
the Sun, caustic ring center and Galactic Center.  It 
relies on a number of idealizations.  How circular the 
ring is is not known.  Near the Sun the actual ring is 
narrower than shown, and its width varies along its 
circumference.  Only the direction of the caustic 
ring center relative to the Sun is known from 
observation.}  
\end{center}
\label{GSCring}
\end{figure}

\begin{figure}
\begin{center}
\includegraphics[height=100mm]{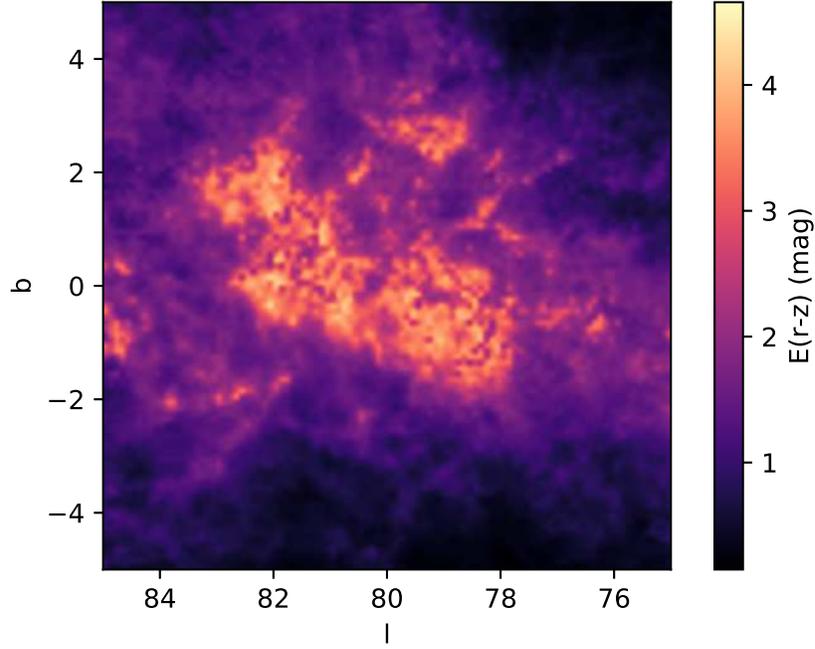} 
\caption{Accumulated reddening from dust at $(l,b) = 
(80^{\circ},0^{\circ})$, up to $3.5\text{ kpc}$. \cite{Green2}}
\end{center}
\label{dust}
\end{figure}

\begin{figure}
\begin{center}
\includegraphics[scale=0.7]{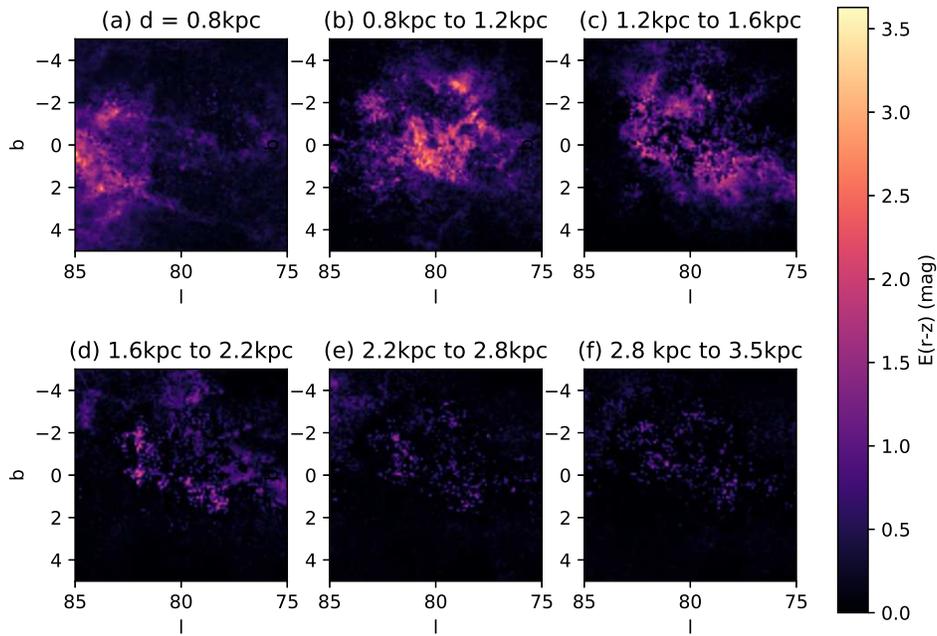}
\vspace{0.3in}
\caption{Accumulated reddening from successive
distance slices at $(l,b) = (80^{\circ},0^{\circ})$. 
The color bar has been rescaled to enhance the 
contrast.}
\end{center}
\label{slice}
\end{figure}

\begin{figure}
\begin{center}
\includegraphics[scale=0.55]{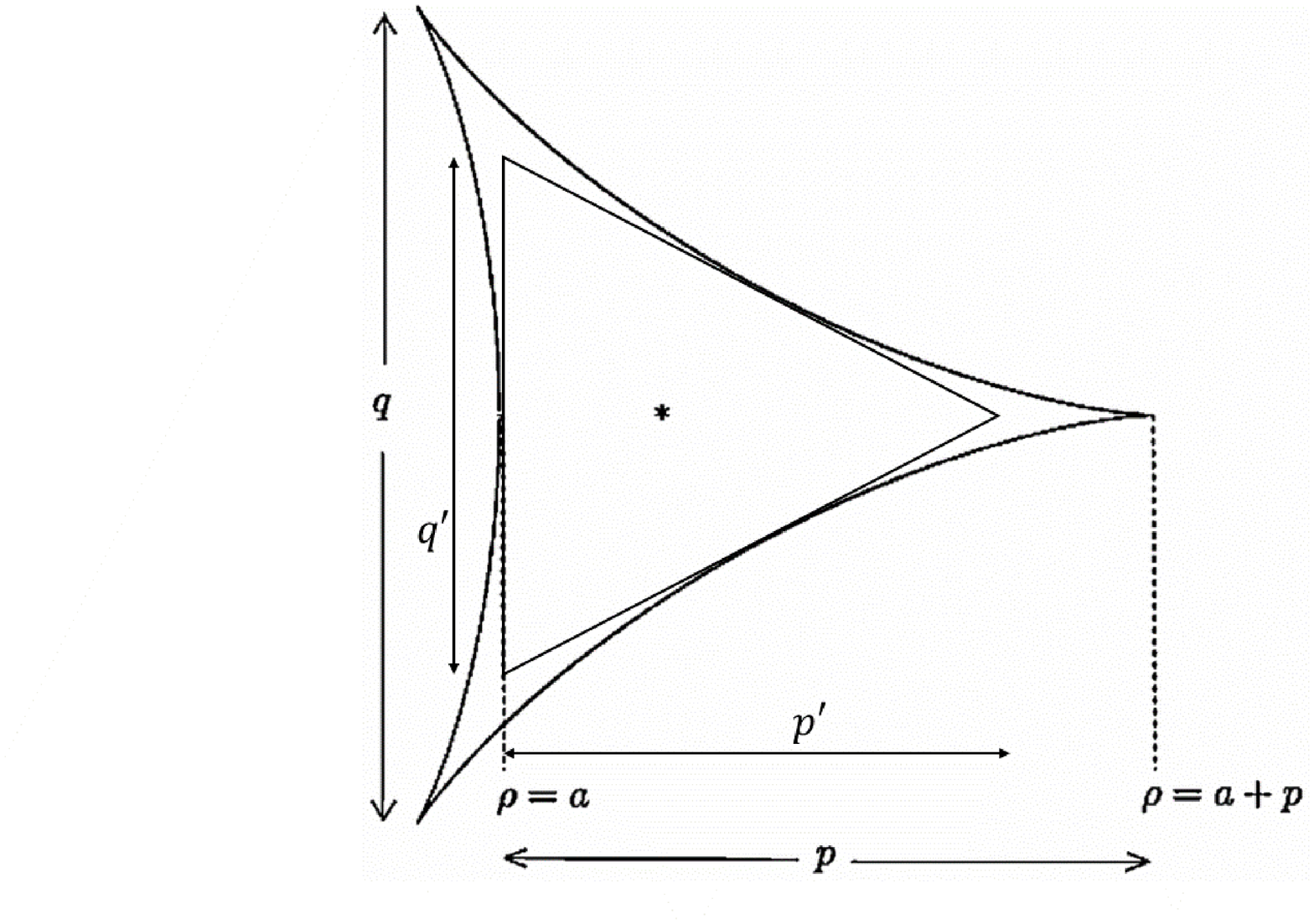}
\vspace{0.3in}
\caption{Triangle inscribed in a tricusp. We 
assume that the observed triangles are in the 
direction of the triangles inscribed in the 5th 
caustic ring tricusp.  The horizontal and vertical 
sizes of a tricusp are related to those of the inscribed 
triangle by: $p = \frac{4}{3} p'$ and $q = \frac{3}{2} q'$.}
\end{center}
\label{inscribed}   
\end{figure}  

\begin{figure}
\begin{center}
\includegraphics[scale=0.5]{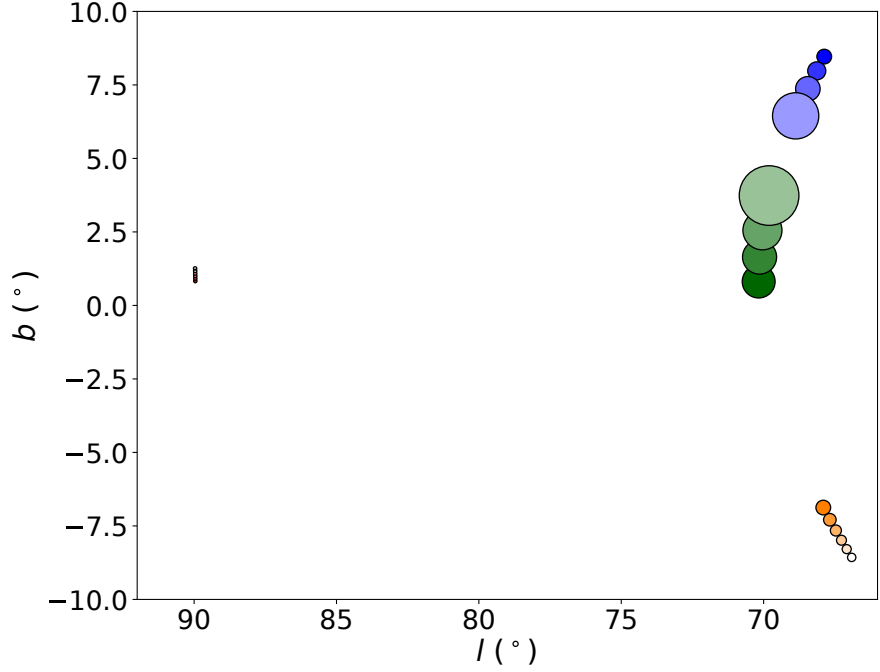}
\vspace{0.3in}
\caption{Galactic Coordinates of the directions of
the four flows associated with the nearby caustic 
ring, at the Sun, with respect to the LSR, in case
$\eta_0 < 0$, for values of the vertical coordinate 
of the Sun $z_\odot$ = 0, 2, 4, 6, 8, 10 pc. The sizes 
of the circles are proportional to the corresponding 
flow densities.  The transparency of the circles 
increases with increasing $z_\odot$. There are 
four flows through the Sun when it is inside the 
tricusp. As the Sun approaches the tricusp boundary, 
two of the flows move towards each other in velocity 
space and their densities increase. When $z_\odot$ 
changes from 6 to 8 pc, the Sun moves outside the 
tricusp and those two flows disappear.} 
\end{center}
\label{varz}   
\end{figure}

\begin{figure}
\begin{center}
\includegraphics[scale=0.65]{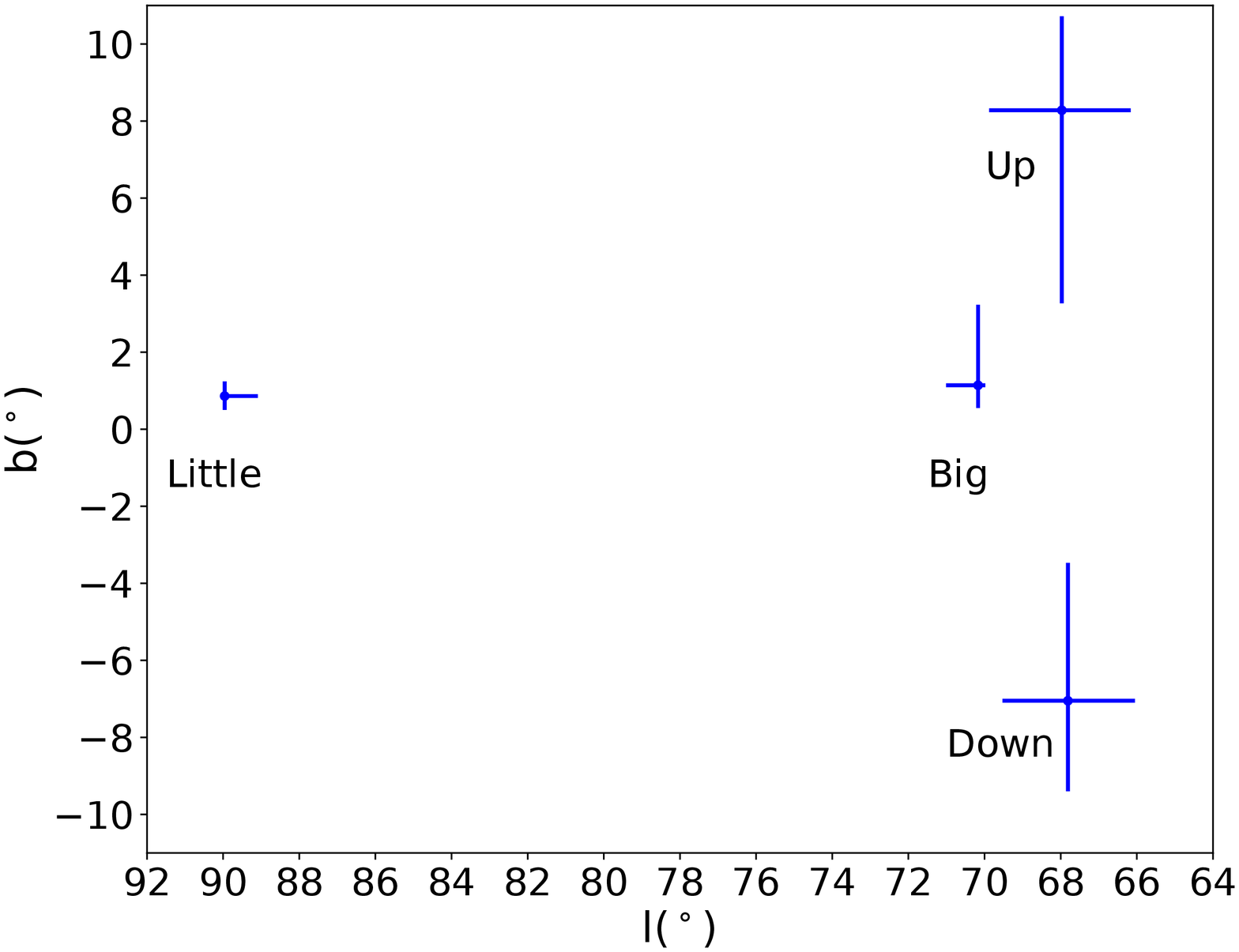}
\vspace{0.3in}
\caption{Galactic Coordinates of the directions of   
the four flows associated with the nearby caustic
ring, at the Sun, with respect to the LSR when $\eta_0 <0$. The error
bars express the uncertainties in our estimates.}
\end{center}
\label{directions}
\end{figure}

\begin{figure}
\begin{center}
\includegraphics[scale=0.65]{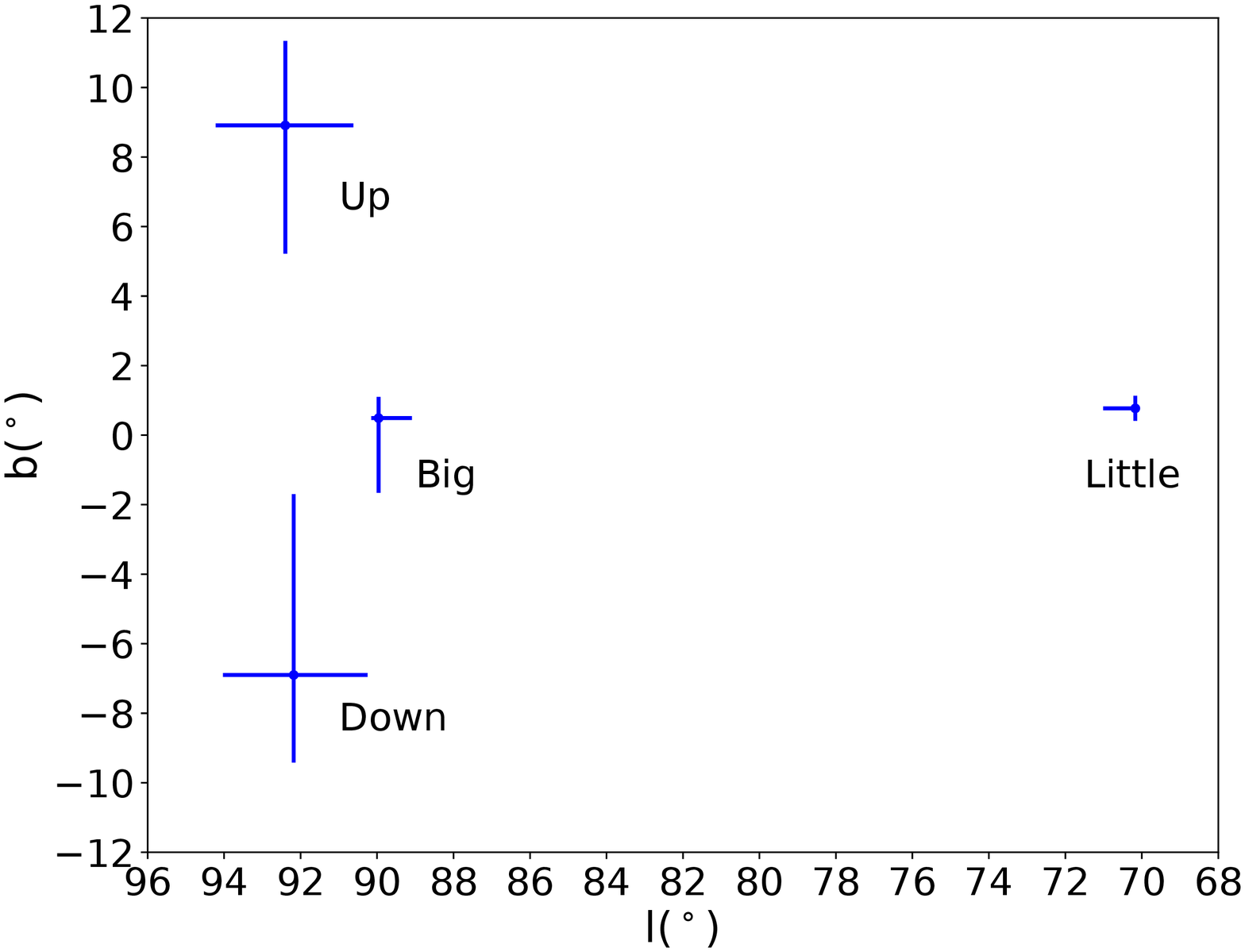}
\vspace{0.3in}
\caption{Galactic Coordinates of the directions of   
the four flows associated with the nearby caustic
ring, at the Sun, with respect to the LSR when $\eta_0 >0$. The error
bars express the uncertainties in our estimates.}
\end{center}
\label{directions2}
\end{figure}

\begin{figure}
\begin{center}
\includegraphics[scale=0.75]{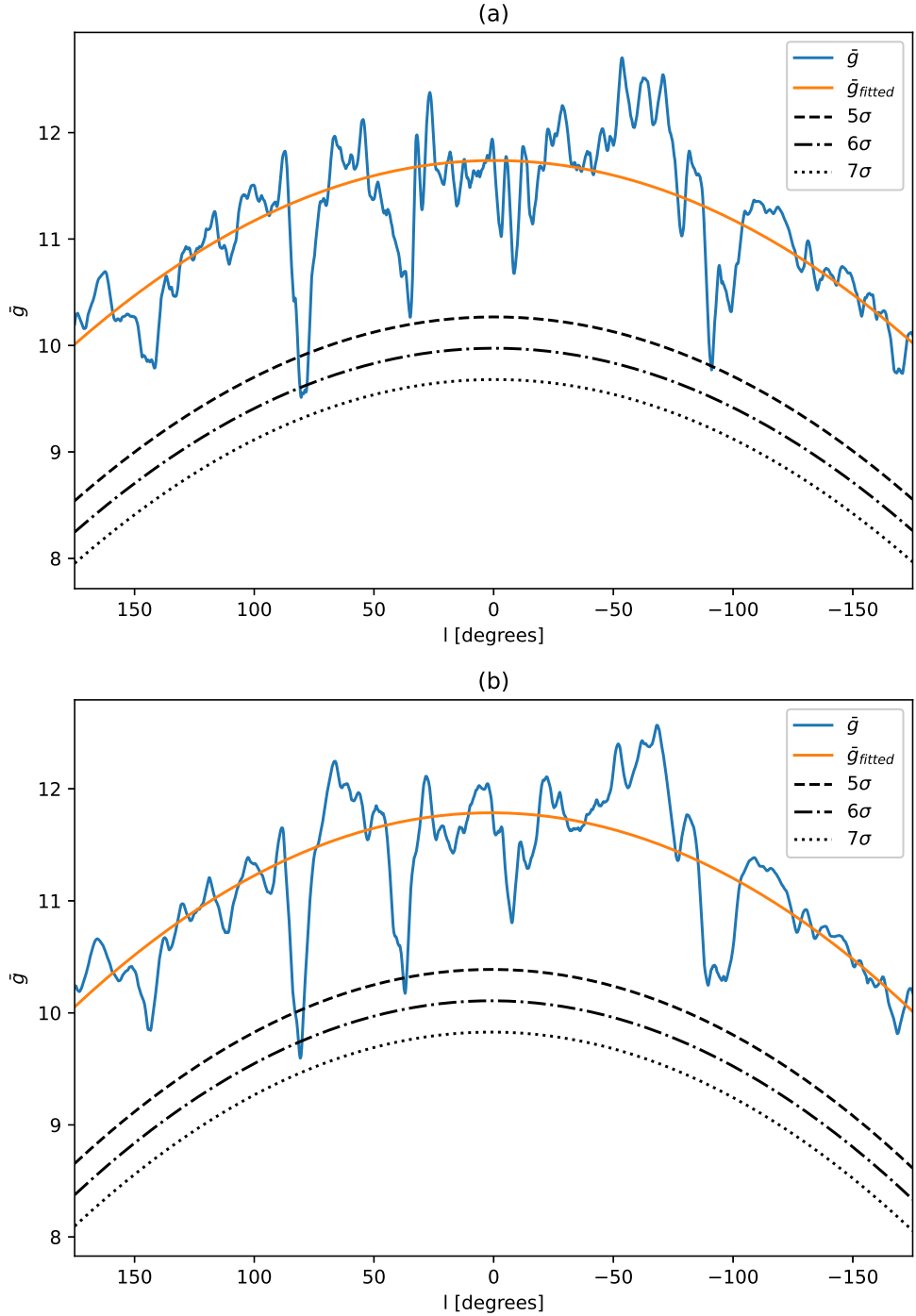}
\vspace{0.3in}
\caption{(a) Convolution of the logarithm of the number of stars 
per square degree in the direction of Galactic Coordinates $(l,b)$
with a triangular top hat filter matching the right triangle in the 
Gaia skymap but displaced in the equatorial direction by an arbitrary 
amount, as a function of its position $l$. (b) Same as in (a) but for 
the left triangle.}  
\end{center}
\label{convoluted}
\end{figure}

\end{document}